\documentclass[reqno]{amsart}
\usepackage{amssymb,cite}
\usepackage{psfig}

\newtheorem{proposition}{Proposition}[section]
\newtheorem{theorem}[proposition]{Theorem}

\newtheorem{lemma}[proposition]{Lemma}

\newcommand{\Rset}{\mathbb{R}}
\newcommand{\Zset}{\mathbb{Z}}

\newcommand{\sech}{\operatorname{sech}}

\newcommand{\cn}{\operatorname{cn}}
\newcommand{\sn}{\operatorname{sn}}
\newcommand{\dn}{\operatorname{dn}}

\newcommand{\lp}{\left(}
\newcommand{\rp}{\right)}
\newcommand{\hDelta}{\hat{\Delta}}

\newcommand{\hk}{\hat{k}}
\newcommand{\hn}{\hat{n}}

\newcommand{\hg}{\hat{g}}

\newcommand{\hT}{\hat{T}}
\newcommand{\ha}{{\hat{a}}}
\newcommand{\hb}{{\hat{b}}}
\newcommand{\hf}{{\hat{f}}}
\newcommand{\hp}{{\hat{p}}}
\newcommand{\hq}{\hat{q}}
\newcommand{\hr}{\hat{r}}
\newcommand{\hv}{\hat{v}}
\newcommand{\cM}{\mathcal{M}}
\newcommand{\cP}{\mathcal{P}}
\newcommand{\cE}{\mathcal{E}}
\newcommand{\hcE}{\hat{\mathcal{E}}}
\newcommand{\cD}{\mathcal{D}}

\begin{document}
\title{Quasi-exact solvability in a general polynomial setting}
\author{David G\'omez-Ullate}
\address{ Dep. F\'isica Te\'orica II, Universidad Complutense de Madrid, 28040 Madrid, Spain. }
\author{ Niky Kamran }
\address{Department of Mathematics and Statistics, McGill University
Montreal, QC, H3A 2K6, Canada}
\author{Robert Milson}
\address{Department of Mathematics and Statistics, Dalhousie University, Halifax, NS, B3H 3J5, Canada} \maketitle

\begin{abstract}
Our goal in this paper is to extend the theory of quasi-exactly
solvable Schr\"odinger operators beyond the Lie-algebraic class. Let
$\cP_n$ be the space of $n$-th degree polynomials in one variable.
We first analyze {\em exceptional polynomial subspaces}
$\cM\subset\cP_n$, which are those proper subspaces of $\cP_n$
invariant under second order differential operators which do not
preserve $\cP_n$. We characterize the only possible exceptional
subspaces of codimension one and we describe the space of second
order differential operators that leave these subspaces invariant.
We then use equivalence under changes of variable and gauge
transformations to achieve a complete classification of these new,
non-Lie algebraic Schr\"odinger operators. As an example, we discuss
a finite gap elliptic potential which does not belong to the
Treibich-Verdier class.

\end{abstract}

\section{Introduction}

Our main focus of study in this paper is the class of second order
linear ordinary differential operators $T$ with analytic
coefficients which preserve a finite dimensional vector space
\[ \mathcal M = \text{span} \{ f_1,\dots,f_n \},\]
where the $f_i(z)$ are polynomials in some coordinate $z$. For a
given $\cM$, we define
\[ \cD_2(\cM)= \{ T\in \cD_2 | T \cM \subset \cM \}\] to be the vector
space of second order differential operators that preserve $\cM$.

The class of operators which belong to the space $\cD_2(\cM)$ for
some $\cM$ is of considerable interest from both a mathematical and
a physical perspective since it encompasses, after a suitable change
of variable, almost all\footnote{We should point out that the
hydrogen atom potential, and more generally, the class of Natanzon
exactly-solvable potentials do not fall directly into this
framework, but that they can be recovered in terms of invariant
spaces of polynomials through a slightly different approach.} the
known examples of potentials for which the Schr\"odinger equation is
either exactly solvable, or quasi-exactly solvable in the sense that
part of the spectrum together with the corresponding eigenfunctions
can be determined algebraically. It is also the basis of much of the
modern theory of orthogonal polynomials.

Historically, the Lie algebra ${\mathfrak{sl}}_{2}(\Rset)$ has
played a rather important role in the development of the subject,
since the vector space ${\mathcal{P}}_{n}$ of univariate
polynomials of degree at most $n$ is an irreducible
${\mathfrak{sl}}_{2}(\Rset)$-module for every positive integer
$n$. This implies that a second-order differential operator which
can be expressed as a polynomial in the generators of
${\mathfrak{sl}}_{2}(\Rset)$ acting on ${\mathcal{P}}_{n}$ will
automatically preserve ${\mathcal{P}}_{n}$. Since the
${\mathfrak{sl}}_{2}$ action on $\cP_n$ is irreducible, Burnside's
Theorem ensures that the converse is also true, i.e. every
operator that preserves $\cP_n$ belongs to the enveloping algebra
of ${\mathfrak{sl}}_{2}$.

 When the operator is put in Schr\"odinger form by a suitable choice of independent
variable and gauge, the resulting potentials, which are referred to
as Lie algebraic potentials, have been widely studied in the
literature. They include the class of P\"oschl-Teller, Morse, sextic
anharmonic, and Lam\'e potentials. One of the limitations of the Lie
algebraic approach is that there is no a-priori reason for it to
encompass all the exactly or quasi-exactly solvable potentials.
Furthermore, it does not provide a systematic way to detect the
presence of multiple algebraic sectors in the spectrum,
corresponding to different subspaces which are preserved by the same
operator.

The general point of our paper is that the class of (quasi-)exactly
solvable potentials is considerably richer than the Lie algebraic
class. We are thus lead to introduce the important concept of an
exceptional subspace $\cM$, defined as a proper subspace of $\cP_n$
 generated by $n$-th degree polynomials such that
$\cD_2(\cM)\not\subset \cD_2(\cP_n)$. Exceptional subspaces lead
naturally to exceptional potentials which are not Lie-algebraic. In
this paper we characterize all exceptional subspaces of co-dimension
one, denoted as $\mathrm X_1$ subspaces (Theorem
\ref{thm:barcelona}), and we provide explicit basis for
$\cD_2(\mathrm X_1)$ (see Proposition \ref{prop:Eop}).  Most
remarkably, both the new $\mathrm X_1$ operators and the Lie
algebraic ones admit an a-priori characterization at the level of
the potential (Theorems \ref{thm:v1111} and \ref{thm:e1111}),
expressed as a linear or a quadratic constraint over the residues
 of the quotient $q(z)/p(z)$ of the coefficients of the
operator $T=p(z) D_{zz}+q(z) D_z + r(z)$. One of the most useful
aspects of this a-priori characterization is that it quickly allows
to decide which potentials are non-singular, and which ones have
multiple algebraic sectors.
Making use of this characterization, and the projective action of
$\mathrm{SL}(2,\Rset)$ on ${\mathcal P}_{n}$, we classify all the
new $\mathrm X_1$ potentials into canonical forms, providing in each
case the potential form in the physical variable with the
constraints over the coefficients and the gauge factor.

As an example of the new potential forms, in Section
\ref{sec:example} we analyze in greater detail a non-singular
elliptic potential on the line which is a deformation of the
Lam\'e potential, admitting four algebraic sectors. This potential
is a finite gap potential which does not belong to the
Treibich-Verdier class \cite{TV90}. The existence of these types
of finite-gap potentials was proved in \cite{TV97} but no results
were given on their solvability. We prove that the potential is
quasi-exactly solvable and we show how to compute the algebraic
eigenfunctions that correspond to the band edges (similar to the
Lam\'e polynomials). These generalized finite-gap potentials have
been studied recently in \cite{Sm06}.

Summarizing, our work illustrates once again the fact that the
representation theory of ${\mathfrak{sl}}_{2}(\Rset)$ is not
needed in order to construct any of the quasi-exactly potentials,
even the Lie algebraic ones, and that reliance on the purely Lie
algebraic setting imposes unnecessary limitations to the theory.

Our paper is organized as follows. In Section \ref{sec:second}, we
recall some basic facts on the transformation properties of
second-order ordinary differential operators, and the equivalence
problem to operators in Schr\"odinger form. Section \ref{sect:sl2}
is devoted to the Lie algebraic class of Schr\"odinger operators and
potentials which arise from finite-dimensional irreducible
representations of ${\mathfrak{sl}}_{2}(\Rset)$ by first order
differential operators.  The classification of Lie-algebraic
potentials reproduces the classifications in \cite{Tu88,GKO93} with
the added feature that potentials with multiple algebraic sectors
are also classified. Section \ref{sect:emod} studies exceptional
subspaces of co-dimension one, and the second order differential
operators that leave them invariant. A characterization of the new
operators at the level of their potential invariant is also given.
The full classification of $\mathrm X_1$ potentials into canonical
forms under the $\mathrm{SL}_2$ projective action is given in
Section \ref{sec:classification}. An $\mathrm X_1$ elliptic finite
gap potential is treated in Section \ref{sec:example}, with an
explicit calculation of its eigenfunctions. Finally Section
\ref{sec:summary} sums up the results and outlines future
developments.


\section{Second order operators.}\label{sec:second}


In this section we consider the general equivalence problem for
second order differential operators in one variable.
%

Let $z$ be a coordinate on a 1-dimensional domain, and let $D_z$
be the unique first order operator such that $D_z[z]=1$. Consider
a general second order operator
\begin{equation}
  \label{eq:Tform}
  T=p(z)D_{zz} + q(z) D_z + r(z),
\end{equation}
where $p,q,r$ are analytic functions of $z$.  We will call
$p(z),q(z)$ and $r(z)$  the \emph{components of $T$ relative to
the coordinate $z$}. We first observe that the components of $T$
are determined by knowing the action of $T$ on three linearly
independent functions.
\begin{proposition}
  \label{prop:Tcomponents}
  Let $f_k(z),\; k=1,2,3$ be three smooth, linearly independent
  functions, and set $g_k(z) = T[f_k(z)]$. The
  components of a second order operator $T$ are the unique solutions
  of the following  linear equation:
  \begin{equation*}
    \begin{pmatrix}
      g_1 \\ g_2 \\ g_3
    \end{pmatrix}=
    \begin{pmatrix}
      f_1'' & f_1' & f_1\\
      f_2'' & f_2' & f_2\\
      f_3'' & f_3' & f_3
    \end{pmatrix}
    \begin{pmatrix}
      p\\q\\r
    \end{pmatrix}\,.
  \end{equation*}
\end{proposition}
The components of a second order differential operator undergo two
very natural transformations: changes of coordinates and gauge
transformations.

A smooth, invertible change of coordinate $z=\zeta(w)$ transforms
the components according to
\begin{align}
  \label{eq:changevar}
  \hp(w) &= \frac{p(\zeta(w))}{\zeta'(w)^2},\\ \nonumber
  \hq(w) &=
  \frac{q(\zeta(w))}{\zeta'(w)}-
  \frac{p(\zeta(w))\zeta''(w)}{\zeta'(w)^3} ,\\ \nonumber
  \hr(w) &= r(\zeta(w)),
\end{align}
where $\hp, \hq, \hr$ denote the components relative to $w$.

We define a  gauge transformation to be the conjugation of an
operator by a multiplication (zero-order) operator:
\begin{equation}\label{eq:gaugetransf}
  \hT = e^{\sigma(z)} T e^{-\sigma(z)}.
\end{equation}
The components of the transformed $\hT$ are given by
\begin{align}
  \label{eq:gaugexform}
  \hp(z)&=p(z),\\ \nonumber
  \hq(z) &= q(z)- 2p(z) \sigma'(z),\\ \nonumber
  \hr(z) &= r(z) - q(z)\sigma'(z)+ p(z)(
  \sigma'(z)^2-\sigma''(z))
\end{align}
We will say that a second order operator $S$ is in
\emph{self-adjoint
  form} if
\begin{equation}
  \label{eq:saform}
  S = p(z) D_{zz} + \frac{1}{2} p'(z) D_z + v(z).
\end{equation}
Every second order operator $T$ can be transformed by a gauge
transformation into self-adjoint form $S$.  Given $T$ with
components $p(z), q(z)$ and $r(z)$ one must take $\sigma(z)$ as
\begin{align*}
  \sigma(z) &= \frac{1}{2}\int^z \frac{q(z)-\frac{1}{2}p'(z)}{p(z)}
  dz.
\end{align*}
In this way, the gauge transformation \eqref{eq:gaugetransf}
transforms $T$
 into self-adjoint form with
\begin{align}
  \nonumber
  v(z) &= r(z) -\frac{1}{2} q'(z)+\frac{1}{4} p''(z)
  -\frac{1}{4p(z)}\lp q(z)-\tfrac{1}{2}p'(z)\rp
  \lp q(z)-\tfrac{3}{2}p'(z)\rp\\
  \label{eq:vform}
  &= r(z) -\frac{1}{2} q'(z)+\frac{1}{4} p''(z)
  -\frac{1}{4}\frac{\lp q-p'(z) \rp^2}{p(z)}+
  \frac{1}{16}\frac{p'(z)^2}{ p(z)}.
\end{align}
Since \eqref{eq:vform} is invariant with respect to gauge
transformations \eqref{eq:gaugexform} we shall refer to $v(z)$ as
the \emph{potential invariant} or simply, the \emph{potential} of
 operator $T$.

We will say that an operator $H$ is in \emph{Schr\"odinger form} if
\begin{equation*}
  H= -D_{xx} + u(x)
\end{equation*}
If $p(z)<0$ on the domain of interest, then an operator $S$ in
self-adjoint form can be transformed into Schr\"odinger form by a
change of coordinate.  The required change of coordinate is the
inverse of the function $x=\xi(z)$, given by
\begin{equation*}
 \xi(z) = \int^z \frac{d z}{\sqrt{-p(z)}}.
\end{equation*}
The potential $u(x)$ is then given by
\begin{equation*}
  u(x) = v(\zeta(x)).
\end{equation*}

Summarizing, an arbitrary second order differential operator $T$
can be transformed into a Schr\"odinger operator by a change of
variable and a gauge transformation. We note that this result is
no longer valid for differential operators in two or more
variables, where the equivalence problem is much harder
\cite{Mil95}. This is the main obstacle to the extension of
quasi-exact solvability to multi-variable operators.

\section{The Lie algebraic class.}
\label{sect:sl2}

As mentioned in the Introduction, our main purpose is to study
differential operators that preserve finite dimensional polynomial
subspaces. More specifically, given a polynomial subspace $\cM$ we
want to give a basis of $\cD_2(\cM)$, the set of second order
differential operators that leave $\cM$ invariant. We then use the
transformations described in the previous Section to construct and
classify Schr\"odinger operators with finite dimensional invariant
subspaces of functions. When the invariant polynomial space is
\begin{equation}\label{eq:Pn}
  \cP_n(z) = \left< 1,z,z^2,\ldots, z^n\right>
\end{equation}
the resulting class of potentials is the well known Lie algebraic
class of quasi-exactly solvable potentials, \cite{Tu88,GKO93}. In
this Section we reproduce the classification of Lie algebraic
potentials using an approach which is different from the usual one
based on the generators of $\mathfrak{sl}_2$ realized as first order
differential operators. Our approach allows to characterize which
 operators are quasi-exactly solvable at the level of the
 potential invariant of the self-adjoint form. The advantage of
 this characterization is that a certain symmetry in the condition
 on the parameters becomes apparent, which allows the same Schr\"odinger operator to be
 algebraized in more than one way, thus leading to
  {\em multiple algebraic sectors}. Examples of this phenomenon had already
  been discussed in the literature \cite{GKM5,KS06,GKM3,FGR00}, but in this Section we provide
  a complete classification of the Lie-algebraic potentials that
  admit multiple algebraizations.
The reason to revisit the well known classification of
Lie-algebraic potentials is that the same approach can be extended
to other polynomial modules, which we deal with in Section
\ref{sect:emod}.

\subsection{ Second order operators that preserve $\cP_n$}

The structure group for the equivalence problem for differential
operators is the infinite-dimensional group generated by
diffeomorphisms and gauge transformations.  If an operator preserves
a finite dimensional vector space of polynomials then the structure
group of the equivalence problem is reduced to $\mathrm{SL}_2$.

The $n+1$ dimensional vector space of polynomials
\begin{equation*}
  \cP_n(z) = \left< 1,z,z^2,\ldots, z^n\right>
\end{equation*}
can be made into an irreducible $\mathrm{SL}_2$ representation. The
irreducible action of $\mathrm{SL}_2$ on $\cP_n$ is given by
\begin{align}
  \label{eq:sl2action}
  \hf(w) &= (\gamma w+\delta)^n f(\zeta(w)),\quad f(z) \in \cP_n(z),\\
  \intertext{where}
  \label{eq:zetaw}
  \zeta(w) &=\frac{\alpha w + \beta}{\gamma
    w+\delta},\quad \alpha \delta - \beta \gamma =1
\end{align}
is a fractional linear transformation corresponding to an element of
$\mathrm{SL}_2$.  The above action is an irreducible multiplier
representation \cite{olver} of $\mathrm{SL}_2$, which is isomorphic
to the unique $n+1$ dimensional irreducible representation of
$\mathrm{SL}_2$. The corresponding infinitesimal generators of the
$\mathfrak{sl}_2$ action are the following first order operators
\begin{equation}
  \label{eq:sl2generators}
  T_- = D_{z},\quad T_0 = z D_{z} - \frac{n}{2},\quad T_+ = z^2
  D_{z} - nz,
\end{equation}
which leave invariant the $n+1$ dimensional space $\cP_n(z)$.

The above $\mathrm{SL}_2$ action  extends naturally to an action on
second order differential operators.  Let $T$ and $\hT$ be second
order differential operators related by the following gauge
transformation:
\begin{equation}
  \label{eq:sl2action-T}
  \hT[\hf(w)] = (\gamma w+\delta)^n T[f(\zeta(w))],
\end{equation}
where $f(z)$ and $\hf(w)$ are related by \eqref{eq:sl2action}, and
 $z=\zeta(w)$ is the fractional linear transformation shown in
\eqref{eq:zetaw}.  Applying \eqref{eq:changevar} and
\eqref{eq:gaugexform}, the transformation law for the operator
components is seen to be
\begin{align}
  \label{eq:sl2action-pqr}
  \hp(w) =&\ (\gamma w+\delta)^4 p(\zeta(w)),\\ \nonumber \hq(w) =&\
  (\gamma w + \delta)^2 q(\zeta(w))-2(n- 1)\gamma(\gamma w + \delta)^3
  p(\zeta(w)), \\ \nonumber \hr(w) =&\ r(\zeta(w))-n\gamma(\gamma w +
  \delta)  q(\zeta(w)) \\
  \nonumber &\ + n(n-1)\gamma^2 (\gamma w + \delta)^2 p(\zeta(w)),
\end{align}
while the transformation law for the corresponding potential
invariants \eqref{eq:vform} is simply
\begin{equation}
  \label{eq:vxform}
  \hv(w) = v(\zeta(w)).
\end{equation}
The inversion transformation $\zeta(w)=-1/w$ plays a significant
role in the subsequent analysis.  We will therefore say that a
second order operator $\hT$ is  the \emph{inversion of operator $T$}
if the two operators are related by
\begin{equation}
  \label{eq:inversion-T}
  \hT[\hf(w)] = w^n T[f(-1/w)],\quad \hf(w) = w^n f(-1/w).
\end{equation}
Specializing \eqref{eq:sl2action-pqr} to the case of an inversion
gives the following transformation rules:
\begin{align}
  \label{eq:inversion-p}
  \hp(w) &= w^4 p(-1/w),\\
  \label{eq:inversion-q}
  \hq(w) &= w^2 q(-1/w) - 2(n-1)w^3 p(-1/w),\\
  \label{eq:inversion-r}
  \hr(w) &= r(-1/w) - nwq(-1/w)+n(n-1)w^2 p(-1/w)
\end{align}
As well, a calculation shows that
\begin{align}
  \label{eq:inversion-p'}
  \hp'(w) &=  w^2 p'(-1/w)+4\, w^3 p(-1/w).
\end{align}

Since $\mathfrak{sl}_2$ acts irreducibly on $\cP_n$, Burnside's
Theorem \cite{Tu88} ensures that a second order operator $T$
preserves $\cP_n$ if and only if it is a quadratic element of the
enveloping algebra of the $\mathfrak{sl}_2$ operators shown in
\eqref{eq:sl2generators}.
 Thus, the most general second order differential operator $T$
 that preserves $\cP_n$ can be written as
\begin{equation}
  \label{eq:liealgebraic}
  T = \sum_{i,j=\pm,\,0} c_{ij} T_i T_j  + \sum_{i=\pm,\,0} b_i T_i+ c^*,
\end{equation}
where $c_{ij}=c_{ji}$, $b_i$ and $c^*$ are constants \footnote{In
the Lie algebraic approach it is more difficult to determine the
dimension of $\cD_2(\cP_n)$ than using the so called direct
approach\cite{GKM1}. The family \eqref{eq:liealgebraic} has {\em a
priori} ten free parameters, but there is one relation among them
coming from the quadratic Casimir of $\mathfrak{sl}_2$, thus
$\dim\cD_2(\cP_n)=9$.}. For this reason, an operator that preserves
$\cP_n(z)$ is often referred to as a \emph{Lie-algebraic} operator.
Note that Burnside's Theorem does not apply to other subspaces $\cM$
generated by polynomials, and therefore an operator $T\in\cD_2(\cM)$
is not necessarily Lie algebraic.

In the rest of the Section we present a different characterization
of operators that preserve $\cP_n$.  Let
\begin{equation*}
  \cP(z) =\bigcup_n \cP_n= \left< 1, z, z^2,\ldots \right>,
\end{equation*}
denote the infinite-dimensional vector space of all polynomials.

\begin{proposition}
  \label{prop:Pops}
  A second order  operator preserves $\cP(z)$ if and only if all its
  components are polynomials in $z$.
\end{proposition}

\begin{proposition}
  \label{prop:allofP}
  If a  second order operator preserves $\cP_n(z)$ for some $n\geq 2$,
  then it preserves $\cP(z)$.
\end{proposition}
\begin{proof}
  Suppose that $T=p(z) D_{zz} + q(z) D_z + r(z)$ preserves
  $\cP_n(z)$. We apply Proposition
  \ref{prop:Tcomponents} with $1,z,z^2$ and obtain
  \begin{equation*}
    r(z) = T[1],\quad
    q(z) = T[z]-z T[1],\quad
    p(z) = \frac{1}{2}\, T[z^2]-z T[z]+\frac{z^2}{2} T[1].
  \end{equation*}
\end{proof}

\begin{proposition}
  \label{prop:liealgT}
  Let $T$ be a  second order differential operator and  $\hT$  its inversion.
  The operator $T$ preserves $\cP_n(z),\; n\geq 2$ if and only if\ $T$
   preserves $\cP(z)$ and $\hT$  preserves $\cP(w)$.
\end{proposition}
\begin{proof}
  If $T$ preserves $\cP_n(z)$, then it preserves $\cP(z)$ by Proposition
  \ref{prop:allofP}. By $\mathrm{SL}_2$ covariance, as expressed by   the
  transformation law
  \eqref{eq:inversion-T}, the transformed operator $\hT$ preserves
  $\cP_n(w)$ and hence $\cP(w)$.

  Conversely, suppose that $T$ and $\hT$ preserve $\cP(z)$ and
  $\cP(w)$, respectively.  Let $f(z)\in\cP_n(z)$ be given and set
  \begin{equation*}
    g(z)=T[f(z)],\quad \hf(w) = w^n f(-1/w)\in\cP_n(w),\quad
    \hg(w)=\hT[\hf(w)].
  \end{equation*}
  By assumption, $g(z)\in\cP(z)$ while $\hg(w)\in\cP(w)$.  However,
  because of covariance
  \begin{equation*}
    \hg(w) = w^n g(-1/w),
  \end{equation*}
  which implies that $\deg g(z)\leq n$. Therefore, $T$ preserves
  $\cP_n(z)$.
\end{proof}

We can now give a basis for $\cD_2(\cP_n)$.  There are other, more
efficient ways to arrive at this result, but we pursue this particular
approach because it will allow us to study $\cD_2(\cM)$ where $\cM$ is
a more general polynomial space.
\begin{proposition}
  \label{prop:labasis}
  If a  second order operator $T=p(z) D_{zz}+q(z) D_z+r(z)$ preserves $\cP_n(z)$, then
  necessarily $\deg p(z)\leq 4$.  Indeed, $T$ preserves $\cP_n(z)$ if
  and only if it is a linear combination of the following 9 operators:
  \begin{align}
    \label{eq:labasis1}
    &z^4 D_{zz} - 2(n-1)z^3 D_z+n(n-1)z^2,\\
    \label{eq:labasis2}
    &z^3 D_{zz} - 2(n-1)z^2 D_z + n(n-1) z,\\
    \label{eq:labasis3}
    &z^2 D_{zz} ,\; z D_{zz},\; D_{zz},\\
    \label{eq:labasis4}
    &z^2 D_z - n z,\; z D_z,\; D_z, 1
  \end{align}
\end{proposition}
\begin{proof}
  Using \eqref{eq:inversion-p}-\eqref{eq:inversion-r} a calculation
  shows that the inversions of the operators shown in
  \eqref{eq:labasis1}-\eqref{eq:labasis4} have polynomial
  coefficients.  Therefore, by Proposition \ref{prop:liealgT}, the
  above 9 operators preserve $\cP_n(z)$.

  Let us now prove that the above operators  span $\cD_2(\cP_n)$.  Let
  $T=p(z)D_{zz}+q(z) D_z + r(z)$ be a second order operator that
  preserves $\cP_n(z)$.  By
  Proposition \ref{prop:liealgT}, the components of the inversion of
  $T$are polynomials in $z$.  An examination of \eqref{eq:inversion-p}
  then shows that, necessarily, $\deg p(z) \leq 4$.  Hence,
  subtracting a linear combination of the operators
  \eqref{eq:labasis1}-\eqref{eq:labasis3}, we obtain a  first order
  operator $T_1=q_1(z)D_z + r_1(z)$ that preserves $\cP_n(z)$.   An
  examination of
  \eqref{eq:inversion-q} shows that, necessarily, $\deg q_1(z)\leq 2$.
  Hence, by subtracting a linear combination of the operators shown in
  \eqref{eq:labasis4}, we arrive at a multiplication (zero-order)
  operator that preserves $\cP_n(z)$.  By \eqref{eq:inversion-r}, such
  an operator can only be a
  constant.
\end{proof}

\subsection{The characterization of Lie-algebraic potentials.}

Following \cite{Ushv94} we recall that the Lie-algebraic character
of an operator can be manifest at the level of its potential
invariant. We shall say that $v(z)$ is a \emph{Lie-algebraic
potential} if there exists a second order Lie-algebraic operator
$T=p(z) D_{zz}+ q(z) D_z + r(z)$ such that $v(z)$ is the
corresponding potential invariant given by \eqref{eq:vform}.

Below we recall the classification \cite{GKO93,Tu88,Ushv94} of
Lie-algebraic potentials based on a case-by-case analysis of the
various configurations of the roots of the fourth degree
polynomial $p(z)$.  We begin by describing the generic case:
\begin{theorem}
  \label{thm:v1111}
  Let $p(z)$ be a 4th degree polynomial with
  distinct roots $z_i$, and let $k_1, k_2, k_3, k_4$ be constants
  satisfying
  \begin{equation}
    \label{eq:sumk}
    k_1+k_2+k_3+k_4=-n-1.
  \end{equation}
  Then,  the rational function
  \begin{equation}
    \label{eq:v1111}
    v(z) = -\sum_{i=1}^4
    \lp k_i{}^2-\tfrac{1}{16}\rp \,\frac{p'(z_i)}{z-z_i}+\lambda.
  \end{equation}
  is a Lie-algebraic potential.  Conversely, let $T=p(z) D_{zz}+q(z)
  D_z + r(z)$ be a second order operator such that $p(z)$ is  a 4th
  degree polynomial with distinct roots $z_i$; let $v(z)$ denote
  the corresponding
  potential, as given by \eqref{eq:vform}; and set
  \begin{equation}
    \label{eq:kidef}
    k_i = \frac{1}{2}\lp \frac{q(z_i)}{p'(z_i)}  -1\rp.
  \end{equation}
  If $T$ preserves $\cP_n(z)$, then
  equations \eqref{eq:sumk} and \eqref{eq:v1111} hold.
\end{theorem}
\begin{proof}
  Suppose that $p(z)$ satisfies the stated assumption, that $k_1, k_2,
  k_3, k_4$ satisfy \eqref{eq:sumk}, and that $v(z)$ has the form
  shown in \eqref{eq:v1111}. Without loss of generality, $p(0)\neq 0$.
  Set $T=p(z)D_{zz} + q(z) D_z + r(z)$, where $q(z)$ is the polynomial
  defined by the equation
  \begin{align}
    \label{eq:qzdef}
    \frac{q(z)}{p(z)}&= \sum_{i=1}^4 \frac{2k_i+1}{z-z_i},
  \end{align}
  and where
    \begin{align*}
    r(z) &= \frac{1}{2} q'(z)-\frac{1}{4} p''(z)
    +\frac{1}{4}\frac{\lp q-p'(z)
      \rp^2}{p(z)}-\frac{1}{16}\frac{p'(z)^2}{ p(z)} \\
  & \qquad - \sum_{i=1}^4
    \lp k_i{}^2-\tfrac{1}{16}\rp \,\frac{p'(z_i)}{z-z_i}+\lambda.
  \end{align*}
  In this way, $v(z)$ is related to the just-constructed operator $T$
  by formula \eqref{eq:vform}.  Taking the limit as $z\to z_i$ in
  \eqref{eq:qzdef} gives \eqref{eq:kidef}.  A calculation shows that
  the residues of $r(z)$ at $z=z_i$ all vanish, and hence $r(z)$ is a
  polynomial.  Therefore, by Proposition \ref{prop:Pops}, $T$
  preserves $\cP(z)$.
  Next, let $\hT=\hp(w) D_{ww} + \hq(w) D_w + \hr(w)$ denote the
  inversion of $T$, and let $\hv(w)=v(-1/w)$ be the corresponding
  potential.  Making the substitutions $z=-1/w,\; z_i = -1/w_i$, and
  using equation \eqref{eq:inversion-p'} we obtain
  \begin{align}
    \label{eq:hvw}
    \hv(w) &= -\sum_{i=1}^4
    \lp k_i{}^2-\tfrac{1}{16}\rp
    \,\frac{\hp'(w_i)}{w-w_i}+\lambda_0,\intertext{where}
    \lambda_0 &= \lambda-\sum_{i=1}^4  \lp
    k_i{}^2-\tfrac{1}{16}\rp \frac{\hp'(w_i)}{w_i}.
  \end{align}
  Applying
  \eqref{eq:inversion-p} and \eqref{eq:inversion-q} gives
  \begin{equation}
    \label{eq:qzqw}
    \frac{\hq(w)}{\hp(w)} = \frac{1}{w^2} \frac{q(-1/w)}{p(-1/w)} +
    (2-2n) \frac{1}{w}.
  \end{equation}
  Equations \eqref{eq:sumk} and \eqref{eq:qzdef} then give
  \begin{align}
    \label{eq:qwdef}
    \frac{\hq(w)}{\hp(w)}&= \lp 2-2n -\sum_{i=1}^4 (2k_i+1)\rp  \frac{1}{w}
      + \sum_{i=1}^4
     \frac{2k_i+1}{w-w_i}\\
     &= \sum_{i=1}^4
      \frac{2k_i+1}{w-w_i}.
  \end{align}
  Hence, mutatis mutandi,  the above argument can be applied to conclude
  that $\hT$ preserves $\cP(w)$, and by Proposition \ref{prop:liealgT}, the
  operator $T$ preserves $\cP_n(z)$.

  Conversely, suppose that $T=p(z)D_{zz}+q(z) D_z + r(z)$ preserves
  $\cP_n(z)$. Let $v(z)$ be defined by \eqref{eq:vform} and $k_i$ be
  the constants defined by \eqref{eq:kidef}.  We denote by
  \begin{align}
    \label{eq:deltaz}
    \Delta(z) &= r(z) -\frac{1}{2} q'(z)+\frac{1}{4} p''(z)
    -\frac{1}{4}\frac{\lp q-p'(z) \rp^2}{p(z)}+
    \frac{1}{16}\frac{p'(z)^2}{ p(z)}\\ \nonumber
    & \qquad  +\sum_{i=1}^4
    \lp k_i{}^2-\tfrac{1}{16}\rp \,\frac{p'(z_i)}{z-z_i}-\lambda
  \end{align}
 the difference of expressions \eqref{eq:vform} and
  \eqref{eq:v1111}.  By Proposition \ref{prop:allofP}, $q(z)$ and $r(z)$
  are polynomials. A calculation shows that $\Delta(z)$ has vanishing
  residues at $z=z_i$, and is therefore a polynomial.  Let $\hT=\hp(w)
  D_{ww}+\hq(w) D_w + \hr(w)$ be the inversion of $T$, and  $\hv(w)
  = v(-1/w)$ its  potential invariant.  Let $\hDelta(w)$ denote
  the difference of $\hv(w)$ and the expression in the right-hand side
  of \eqref{eq:hvw}.  Repeating the above argument, we conclude that
  $\hDelta(w)$ is a polynomial in $w$. Making the substitutions
  $z=-1/w, z_i=-1/w_i$ in \eqref{eq:deltaz} and using
  \eqref{eq:inversion-p'} we obtain that $\hDelta(w) =
  \Delta(-1/z)$, which
  is only possible if both sides are constant.  Hence, $v(z)$ has
  the form, up to a constant term, shown in \eqref{eq:v1111}.  Since
  both $q(z)$ and $\hq(w)$ are polynomials, equation \eqref{eq:qzqw}
  implies that $\deg q \leq 3$.  Hence, equation \eqref{eq:kidef}
  implies equation \eqref{eq:qzdef} and  since $\hq(w)$ is a
  polynomial, then
  equation \eqref{eq:qwdef} implies \eqref{eq:sumk}.
\end{proof}

\subsection{The generalized Lam\'e potentials.}
Suppose that $p(z)$ has four distinct roots.  In describing the
generic Lie algebraic potential, no generality is lost if we
assume that one of the roots is at infinity, i.e., that $\deg
p=3$.  One can further specialize the form of the above potentials
by stipulating
\begin{equation*}
  z_1=1-\frac{1}{m},\quad z_2=0,\quad z_3=1,\quad   z_4=\infty,\quad 0<m<1.
\end{equation*}
Specializing Theorem \ref{thm:v1111}, we obtain the following.
\begin{proposition}
  \label{prop:v111}
  Let $S$ be an operator in self-adjoint form \eqref{eq:saform}, and
  suppose that $p(z)=4z(1-z)(mz-m+1)$.  The corresponding
  potential $v(z)$ is Lie-algebraic if and only if
  \begin{align}
    \label{eq:v111}
    v(z) =&\  4\,(k_1{}^2-\tfrac{1}{16})\,m\, z-
    (k_2{}^2-\tfrac{1}{16})\,\frac{(m-1)}{mz-m+1}+\\ \nonumber
    &\ +
    (k_3{}^2-\tfrac{1}{16})\,\frac{(m-1)}{z} +
    (k_4{}^2-\tfrac{1}{16})\,\frac{1}{z-1} ,
  \end{align}
  such that
  \begin{equation}
    \label{eq:sumk1}
    k_1+k_2+k_3+k_4=-n-1.
  \end{equation}
\end{proposition}
\subsection{Multiple algebraic sectors.}
It has long been known that some exactly solvable potentials possess
multiple algebraic sectors.  For example, the harmonic oscillator
possess one algebraic sector corresponding to even eigenfunctions,
and another corresponding to odd eigenfunctions.  The existence of
multiple algebraizations for the Lam\'e potential has been noted in
\cite{brihaye03,FGR00}

We will say that a second-order operator $T$ possesses multiple
$\mathrm{SL}_2$ algebraizations if there exists a local coordinate
$z$ such that $T$ preserves both the vector space $\cP_{n}(z)$ and
the vector space $\phi(z) \cP_{\hn}(z)$.  A distinct, second
algebraic sector arises if $\phi(z)$ is not a polynomial, leading to
more algebraic eigenfunctions.  In some cases, when $\phi(z)$ is a
polynomial, it happens that $\phi(z) \cP_{\hn}(z)\subset \cP_n(z)$
and no new algebraic eigenfunctions arise.  We say then that the
operator $T$ possesses \emph{nested} $\mathrm{SL}_2$
algebraizations.

Multiple $\mathrm{SL}_2$ algebraizations have been discussed in
\cite{brihaye03,FGR00,GKM3}, but a full classification of potentials
admitting multiple $\mathrm{SL}_2$ algebraizations was not known.
Here we provide the full classification thanks to the explicit
potential form shown in equations \eqref{eq:sumk}-\eqref{eq:v1111}.
  In the generic case, when
$p(z)$ has simple roots, multiple $\mathrm{SL}_2$ algebraizations
arise from the parameter symmetry
\begin{equation*}
  \hk_i = (-1)^{p_i} k_i,\quad \hn =n +
  \sum_{i=1}^4(1-(-1)^{p_i})k_i,\quad
  p_i \in \{ 0,1 \}.
\end{equation*}
in the formula for $v(z)$ shown in \eqref{eq:v1111}.
The possible parameter symmetries are given by the following
\begin{proposition}
  Let $n$ be an integer.  The above $\hn$ is an integer if and only if
  $2k_i$ is an integer for some $i$, or $2(k_i\pm k_j)$ is an integer for
  some $i\neq j$.
\end{proposition}
\noindent Consequently, there are 4 cases in which multiple algebraic
sectors arise.
\begin{itemize}
\item[(M2a)] If $k_i\in \tfrac{1}{2}\,\,\Zset$ for exactly one $i$, there
are two algebraic sectors.
\item[(M2b)] If there is no $i$ such that $k_i\in \tfrac{1}{2}\,\Zset$,
  but $k_i\pm k_j\in \tfrac{1}{2}\,\Zset$ for exactly two
  choices of $i\neq j$, then there
  are two algebraic sectors.
\item[(M4a)] If $k_i\in \tfrac{1}{2}\,\Zset$ for exactly two values of
  $i$, there are four algebraic sectors.
\item[(M4b)] If $k_i\in \tfrac{1}{4}\,\Zset$ for all $i$, there are
  four algebraic sectors.
\item[(M8)] If $k_i\in \tfrac{1}{2}\,\Zset$ for all $i$, there are eight algebraic sectors.
\end{itemize}

We observe that if the root structure of $p(z)$ is degenerate, i.e.
if $\deg p(z)\leq 2$ the opportunity for multiple algebraizations is
lessened, but some possibilities still remain. The full
classification of multiple $\mathrm{SL}_2$ algebraizations is given
in Tables \ref{tab:multalgI}-\ref{tab:multalgD}.  As above, the
classification is achieved by examining all the possible symmetries
in the parameters of the potential.


Note that here we restrict ourselves to multiple $\mathrm{SL}_2$
algebraizations, in which the invariant polynomial spaces are of the
same type. However, other multiple algebraizations exist \cite{GKM4}
in which the invariant polynomial spaces are of a different type,
namely $\cP_n$ and an exceptional polynomial space, like those
described in Section \ref{sect:emod}.

\subsection{Classification of Lie-algebraic potentials}

We finally describe the classification of Lie-algebraic potentials.
The novelty with respect to previous classifications is that the
potentials which admit multiple algebraic factors have also been
classified.

We will refer\footnote{We borrow the nomenclature for the root
structure of
  4th degree polynomial from the Penrose-Petrov classification of the
  4-dimensional Weyl tensor.} to the potentials in proposition
\ref{prop:v111} as type I .  If $p(z)$ has a multiple root, no
generality will be lost if we assume that this root is located at
infinity.  Normalizing the value of the other root(s) to various
convenient values, and taking limits of \eqref{eq:v111} and
\eqref{eq:sumk1}, as appropriate, we obtain the remaining cases.

\subsubsection{Type I potentials.}
\begin{align}
  \nonumber
  p(z) = &\ 4z(1-z)(mz-m+1),\quad   z=\cn(x|m)^2,\quad m\in (0,1),\\
  v(z) =& \ 4\,(k_1{}^2-\tfrac{1}{16})\,m\, z-
  (k_2{}^2-\tfrac{1}{16})\,\frac{(m-1)}{mz-m+1}+\\ \nonumber &\  +
  (k_3{}^2-\tfrac{1}{16})\,\frac{(m-1)}{z} +
  (k_4{}^2-\tfrac{1}{16})\,\frac{1}{z-1}  ,\\
  \nonumber n =&-(k_1+k_2+k_3+k_4)-1,\\ \nonumber
  \mu(x) =& \dn(x|m)^{c_2}
  \cn(x|m)^{c_3} \sn(x|m)^{c_4}, \quad c_i = 2k_i+1/2.
\end{align}
Taking $k_3, k_4=\pm 1/4$, makes the potential non-singular on the
interval $0\leq z\leq 1$.  The resulting class is referred to as the
generalized or the associated Lam\'e potentials \cite{GKM3}.  The
ordinary Lam\'e potentials are obtained by setting $k_2=\pm 1/4$.
The possibilities for multiple algebraizations are indicated in Table
\ref{tab:multalgI}.  Requiring that the  potential be  non-singular
eliminates many of the possibilities.
\begin{table}[htbp]
  \centering
  \small
  \begin{tabular}{@{\vbox to 12pt{}}clc}
    \hline
    \hfil $\phi$ \hfil & \hfil $\hn-n$\hfil &\hfil Conditions\hfil \\ \hline
    $(z-1+\frac{1}{m})^{-2k_2} z^{-2k_3} (z-1)^{-2k_4}$
    & $2(k_2+k_3+k_4)$ &
    $\vert k_1 \vert  > \vert k_2+k_3+k_4 \vert$ \\
    $z^{-2k_3} (z-1)^{-2k_4}$
    & $2(k_3+k_4)$ &
    $\vert k_1+k_2 \vert  > \vert k_3+k_4 \vert$ \\
    $\ldots$  &  $2 (k_1+k_3+k_4)$ & $ | k_2 | > |k_1+k_3+k_4|$\\
    $ (z-1+\frac{1}{m})^{-2k_2} z^{-2k_3}$
    & $2(k_2+k_3)$ &
    $\vert k_1+k_4 \vert  > \vert k_2+k_3 \vert$ \\
    $\ldots$  &  $2 (k_1+k_2+k_3)$ & $ | k_4 | > |k_1+k_2+k_3|$\\
    $ (z-1+\frac{1}{m})^{-2k_2} (z-1)^{-2k_4}$
    & $2(k_2+k_4)$ &
    $\vert k_1+k_3 \vert  > \vert k_2+k_4 \vert$ \\
    $\ldots$  &  $2 (k_1+k_2+k_4)$ & $ | k_3 | > |k_1+k_2+k_4|$\\

    $ (z-1+\frac{1}{m})^{-2k_2}$
    & $2k_2$ &
    $\vert k_1+k_3+k_4 \vert  > \vert k_2 \vert$ \\
    $\ldots$  &  $2 (k_1+k_2)$ & $ | k_3+k_4 | > |k_1+k_2|$\\

    $ z^{-2k_2}$
    & $2k_3$ &
    $\vert k_1+k_2+k_4 \vert  > \vert k_3 \vert$ \\
    $\ldots$  &  $2 (k_1+k_3)$ & $ | k_2+k_4 | > |k_1+k_3|$\\

    $ (z-1)^{-2k_4}$
    & $2k_4$ &
    $\vert k_1+k_2+k_3 \vert  > \vert k_4 \vert$ \\
    $\ldots$  &  $2 (k_1+k_4)$ & $ | k_2+k_3 | > |k_1+k_4|$\\

    $ 1 $
    & $2k_1$ &
    $\vert k_2+k_3+k_4 \vert  > \vert k_1 \vert$ \\ \hline \\
  \end{tabular}
  \caption{Multiple algebraizations for type I potentials}
  \label{tab:multalgI}
\end{table}

It is worth mentioning that the class of type I Lie algebraic
potentials described above coincides with the well known class of
Treibich-Verdier elliptic finite gap potentials,
\cite{KS06,TV90,Tak04b}.

\subsubsection{Type II potentials.}
\begin{align}
  \nonumber
  &p(z) = 4z(1-z),\quad z=-\sinh^2 x,\\
  &v(z) = 4\,
  \ell_1{}^2\,z^2 -4\, \ell_1(\ell_1+2\ell_2)\, z-
  \frac{k_3{}^2-\tfrac{1}{16}}{z}-\frac{k_4{}^2-\tfrac{1}{16}}{1-z} ,\\
  \nonumber
  &n=-(\ell_2+k_3+k_4)-1,\\ \nonumber
  & \mu(x) = \exp(-\ell_1\sinh^2 x) (\sinh x)^{c_3} (\cosh
  x)^{c_4},\quad c_i=2k_i+\frac{1}{2}.
\end{align}
There are multiple algebraic sectors if one of the three
parameters $\ell_2, k_3, k_4$ is a half-integer.  If all three
parameters are half-integers then there are four algebraic sectors
(possibly nested).

The non-singular potentials arise when $k_3=\pm 1/4$, and in this
case the sign of $k_3$ governs the parity of the eigenfunctions.
Thus, there is one type of non-singular type II potentials with
multiple algebraic sectors.  (line 1 of Table
\ref{tab:multalgII}).  

\begin{table}[htbp]
  \centering
  \begin{tabular}{@{\vbox to 12pt{}}clc}
    \hline
    \hfil $\phi$ \hfil & \hfil $\hn-n$\hfil &\hfil Conditions\hfil \\
    \hline
    $z^{-2k_3} (z-1)^{-2k_4}$ & $2(k_3+k_4)$ & $|\ell_2| >
    |k_3+k_4|$\\
    $z^{-2k_3}\, e^{-2 \ell_1 z}$ & $2(\ell_2+k_3)$ & $|k_4| >
    |\ell_2+k_3|$ \\
    $(z-1)^{-2k_4}\, e^{-2 \ell_1 z}$ & $2(\ell_2+k_4)$ & $|k_3| >
    |\ell_2+k_4|$ \\
    $e^{-2 \ell_1 z}$ & $2\ell_2$ & $|k_3+k_4| >
    |\ell_2|$ \\
    $z^{-2k_3}$ & $2k_3$ & $|\ell_2+k_4| >
    |k_3|$ \\
    $(z-1)^{-2k_4}$ & $2k_4$ & $|\ell_2+k_3| >
    |k_4|$ \\ \hline \\
  \end{tabular}
  \caption{Multiple algebraizations for type II potentials}
  \label{tab:multalgII}
\end{table}
\subsubsection{Type D potentials}
\begin{align}
  \nonumber
  & p(z) = -z^2, \quad z=\exp(x),\\
  & v(z) = \ell_1{}^2 z^2 -2\, \ell_1 \ell_2 \,z + \ell_3{}^2 z^{-2}
  +2\, \ell_3\ell_4 \,z^{-1},\\ \nonumber
  & n=-(\ell_2+\ell_4)-1,\\ \nonumber
  & \mu(x) = \exp(\ell_1 e^x - \ell_3\,
  e^{-x}+\ell_4\,x).
\end{align}
These potentials are never singular.  There are two algebraic sectors
whenever $\ell_2,\ell_4\in\frac{1}{2}\, \Zset$.
\begin{table}[htbp]
  \centering
  \begin{tabular}{@{\vbox to 14pt{}}clc}
    \hline
    \hfil $\phi$ \hfil & \hfil $\hn-n$\hfil &\hfil Conditions\hfil \\
    \hline
    $e^{(2\ell_3)/z} \,z^{-2 \ell_4}$ & $2 \ell_4$&  $|\ell_2| >
    |\ell_4|$ \\
    $e^{2\ell_1 z}$ & $2 \ell_2$&  $|\ell_4| > |\ell_2|$ \\
    \hline \\
  \end{tabular}
  \caption{Multiple algebraizations for type D potentials}
  \label{tab:multalgD}
\end{table}

\subsubsection{Type Z potentials}
\begin{align}
  & p(z) = -(z^2+1),\quad z=\sinh x\\ \nonumber & v(z) = \ell_1{}^2\,
  z^2 -2 \, \ell_1\ell_2\, z+\frac{2 \ell_3\ell_4 z
    +\ell_3{}^2-\ell_4{}^2+1/4}{z^2+1},\\ \nonumber &
  n=-(\ell_2+\ell_4)-1,\\ \nonumber & \mu(x) = \exp\left\{\ell_1 \,
    \sinh x +\ell_3
    \tan^{-1}(\sinh x)\right\} (\cosh x)^{\ell_4}.
\end{align}
These potentials are never singular.  There are 2 algebraic sectors if
$\ell_2,\ell_4$ are half-integers.
\begin{table}[htbp]
  \centering
  \begin{tabular}{@{\vbox to 14pt{}}clc}
    \hline
    \hfil $\phi$ \hfil & \hfil $\hn-n$\hfil &\hfil Conditions\hfil \\
    \hline
    $e^{-2\ell_1 z}$ & $2 \ell_2$&  $|\ell_4| >
    |\ell_2|$ \\
    $e^{-2\ell_3 \tan^{-1}(z)}\, (z^2+1)^{-\ell_4}$ & $2 \ell_4$&
    $|\ell_2| > |\ell_4|$ \\
    \hline \\
  \end{tabular}
  \caption{Multiple algebraizations for type Z potentials}
  \label{tab:multalgD}
\end{table}

\subsubsection{Type III potentials}
\begin{align}
  & p(z) = -4z ,\quad z=x^2,\\ \nonumber
  & v(z) = \ell_3{}^2 z^3 -4\, \ell_2
  \ell_3\, z^2+4\, (\ell_2{}^2+\ell_1\ell_3) z+ \lp
  k_4{}^2-\tfrac{1}{16}\rp z^{-1} , \\
  \nonumber & n=-(\ell_1+k_4)-1,\\ \nonumber
  &\mu(x) = \exp\left\{-\tfrac{1}{4}\ell_3 x^4 + \ell_2 x^2\right\}
x^{c_4},\quad c_4 = 2k_4+\frac{1}{2}.
\end{align}
These are the well-known QES sextics.  They are non-singular if
and only if $k_4=\pm 1/4$, in which case the sign of $k_4$ governs
the parity of the eigenfunctions.  There are two algebraic sectors
if and only if $\ell_1, k_4 \in \frac{1}{2}\, \Zset$.  Thus, a
non-singular QES sextic possesses only one Lie-algebraic sector.
\begin{table}[htbp]
  \centering
  \begin{tabular}{@{\vbox to 14pt{}}clc}
    \hline
    \hfil $\phi$ \hfil & \hfil $\hn-n$\hfil &\hfil Conditions\hfil \\
    \hline
    $e^{\frac{1}{2}\ell_3 z^2-2\ell_2 z}$ & $2 \ell_1$&  $|k_4| >
    |\ell_1|$ \\
    $z^{-2k_4}$ & $2 k_4$&
    $|\ell_1| > | k_4|$ \\
    \hline \\
  \end{tabular}
  \caption{Multiple algebraizations for type III potentials}
  \label{tab:multalgD}
\end{table}
\subsubsection{Type N potentials}
\begin{align}
  & p(z) = -1 ,\quad z=x,\\ \nonumber
  & v(z) = (\ell_1 z +\ell_2)^2 z^2 +(\ell_1 z + \ell_2)(\ell_3 z + \ell_4),\\
  \nonumber
  &\ell_4 = -2-2n,\\ \nonumber
  & \mu(x) = \exp\left\{\tfrac{1}{3} \ell_1 x^3-\tfrac{1}{2}\ell_2
    x^2-\tfrac{1}{2} \ell_3x\right\}
\end{align}
To obtain normalizable eigenfunctions, one must take
$\ell_1=0,\ell_2>0$. The result is a harmonic oscillator
potential.
\section{Exceptional subspaces.}
\label{sect:emod}

Let $\cM\subset \cP_n$ be a finite dimensional subspace generated by
polynomials.  We define $\cM$ to be an {\em exceptional subspace of
$\cP_n$} if $\cD_2(\cM)\not\subset \cD_2(\cP_n)$. The significance
of exceptional subspaces is clearly that they are the only ones that
lead to quasi-exactly solvable potentials which do not belong to the
Lie-algebraic class. For brevity we will denote by $\mathrm{X}_k$ an
{\em exceptional subspace of co-dimension $k$}. In the rest of the
paper we study and classify all QES potentials whose invariant
subspace is $\mathrm{X}_1$.

The analysis \cite{PT95, GKM3} of polynomial subspaces spanned by
monomials brought to light two special subspaces:
\begin{align}\label{eq:emod}
  \cE_n(z) & =
  \left<1,z^2,\dots,z^n\right>, \\ \label{eq:emod2}
  \hcE_n(z) &=
  \left< 1,z,z^2,\dots,z^{n-2},z^n\right>.
\end{align}
These two subspaces $\cE_n$ and $\hcE_n$ are
$\mathrm{SL}_2$-equivalent, since
\begin{equation*}
  \hcE_n(w) = w^n \cE_n(-1/w) = \{ w^n f(-1/w) \colon f(z)\in\cE_n(z)\}.
\end{equation*}


The following result shows that $\cE_n$ and $\hcE_n$ are not only
special within the class of spaces generated by monomials, but that
they are essentially the only $\mathrm{X}_1$ subspaces.
\begin{theorem}\label{thm:barcelona}
  Let $\cM\subset\cP_n$ be an $n$-dimensional polynomial subspace (not
  necessarily spanned by monomials), and suppose that $\cM\neq
  \cP_{n-1}$.  Then, $\cM$ is  an
  exceptional subspace if and only if $\cM$ is $\mathrm{SL}_2$-equivalent
  to $\cE_n$.
\end{theorem}
\noindent This theorem limits the only interesting codimension-one
polynomial subspaces for which new QES potentials can be
constructed. The proof is rather lengthy and shall be given
elsewhere.

We devote the rest of the paper to the classification of the new
$\mathrm{X}_1$ potentials.
For this purpose, we need to give a basis of $\cD_2(\cE_n)$ and then
describe the potentials that correspond via \eqref{eq:vform} to such
operators. First, we inquire about the effect of an $\mathrm{SL}_2$
transformation on $\cE_n$. The effect of an inversion
transformation, $z=-1/w$, was described above. More generally, let
us set
\begin{align}
  \cE(z)&=\left< 1,z^2,z^3,\ldots \right>,\\
  \label{eq:Ebasis2}
  \cE^{a,b}(z) &= \left < a(z-b)-1,(z-b)^2,(z-b)^3,\ldots \right>,\\
  \label{eq:Ebasis1}
  \cE_n^{a,b}(z) &= \left< a(z - b)-1,(z-b)^2,(z-b)^3, \ldots,
  (z-b)^n\right>,\\  \nonumber
  \hcE_n^{a}(z) &=   \left< 1,z,z^2,\ldots ,z^{n-2},
  z^n  -  a z^{n-1}\right>,
\end{align}
and note that $\cE_n^{a,b} = \cE^{a,b}\cap \cP_n$.

\begin{proposition}
  \label{prop:Ebasis}
  The infinite-dimensional vector space $\cE^{a,b}$ consists of
  polynomials $f\in \cP$ that satisfy by the following first order
 constraint:
  \begin{equation}
    \label{eq:econd}
    f'(b) + a f(b) = 0
  \end{equation}
\end{proposition}
\noindent
\begin{proposition}
  \label{prop:Ebasis1}
  The subspaces $\cE^{a,b}_n$ and $\hcE^a_n$ are $\mathrm{X}_1$. To be more
  precise, let
  $$z=\zeta(w)=(\alpha w + \beta)/(\gamma w+\delta),\quad \alpha
  \delta - \beta \gamma =1$$
  be a fractional linear transformation. If
  $\alpha\neq 0$, we have
  \begin{equation}
    \label{eq:Ebasis3}
    (\gamma w+\delta)^n \cE_n(\zeta(w)) =\cE_n^{a,b}(w),
  \end{equation}
  where $a = - n\alpha\gamma, b = -\beta/\alpha$.
  If $\alpha=0$, then
  \begin{equation}
    \label{eq:Ebasis4}
    (\gamma w+\delta)^n \cE_n(\zeta(w)) =\hcE_n^{a}(w),
   \end{equation}
   where $a=-n\delta/\gamma$.
\end{proposition}
\begin{proof}
In the
  first case, using $\alpha\delta-\beta\gamma=1$, we have
  \begin{align*}
    (\gamma w+\delta)^n \cE_n(\zeta(w)) &= \left<(\gamma w +
      \delta)^{n-k} ( \alpha w + \beta)^k
      \,\colon\,k=0,2,\dots,n\right> \\
    &= \left<(-a(w-b) + n)^{n-k} ( w -b)^k
      \,\colon\,k=0,2,\dots,n\right>,\\
    &= \cE^{a,b}_n(w).
  \end{align*}
  The last equality is true because $(-a(w-b) + n)^{n-k}( w -b)^k$
  satisfies \eqref{eq:econd} for $k=0$ and for $k\geq 2$.
  If $\alpha=0$, then
  \begin{align*}
    (\gamma w+\delta)^n \cE_n(\zeta(w)) &= \left<
      (w-a/n)^{n-k}\colon\,k=0,2,\dots,n\right> \\
    &= \left<1,w,w^2,\ldots,w^{n-2},w^n-a w^{n-1}\right>\\
    &= \hcE^{a}_n(w)
  \end{align*}
Since $\cE^{a,b}_n$ and $\hcE^a_n$ are $\mathrm{SL}_2$ equivalent to
$\cE_n$, by Theorem \ref{thm:barcelona} they are $\mathrm{X}_1$.
\end{proof}
We  describe explicitly the relationship between exceptional
subspaces related by an inversion $w=-1/z$:
\begin{proposition}
  \label{prop:Ebasis2}
  For $b\neq 0$, the $\mathrm X_1$ subspaces are related by an inversion
  \begin{equation}
    \label{eq:Ebasis2}
    u^n \cE^{a,b}_n(-1/u)=\cE^{\ha,\hb}_n(u),\quad\text{where}\quad\ha
    = b(n+ab),\quad\hb=-1/b.
  \end{equation}
  For $b=0$, we have
  \begin{equation}
    \label{eq:Ebasis2a}
    u^n \cE^{a,0}_n(-1/u)=\hcE^{-a}_n(u).
  \end{equation}

\end{proposition}
\begin{proof}
  Let $z=(\alpha w + \beta)/(\gamma w + \delta),\;
  \alpha\delta-\beta\gamma=1$ be a fractional
  linear transformation.  Suppose that $\alpha,\beta\neq 0$.
  With $w=-1/u$, we have $z=(\beta u
  -\alpha)/(\delta u - \gamma)$.  Setting
  \begin{equation*}
    b=-\beta/\alpha,\quad  a=-n\alpha\gamma,\quad
    \hb=\alpha/\beta,\quad \ha = -n\beta\delta,
  \end{equation*}
  we have, by Proposition
  \ref{prop:Ebasis1},
  \begin{align*}
    &(\gamma w+\delta)^n \cE_n(z) =\cE_n^{a,b}(w),\\
    &(\delta u-\gamma)^n \cE_n(z) =\cE_n^{\ha,\hb}(u).
  \end{align*}
  From this, \eqref{eq:Ebasis2} follows.

  To prove \eqref{eq:Ebasis2a}, suppose that $z, w, u$ are as above,
  but that $\beta=0$.  Since $n\gamma/\delta=n\alpha\gamma=-a$, we
  have
  \begin{align*}
    &(\gamma w+\delta)^n \cE_n(z) =\cE_n^{a,0}(w),\\
    &(\delta u-\gamma)^n \cE_n(z) =\hcE_n^{-a}(u).
  \end{align*}
\end{proof}
Our main interest in this Section is to study second order operators
$T$ that preserve $\cE^{a,b}_n(z)$ but {\em do not preserve
$\cP_n(z)$}, which we shall denote as \emph{$\mathrm{X}_1$
operators}. These are the only non-Lie algebraic operators (up to
$\mathrm{SL}_2$ equivalence) that preserve a codimension one
polynomial subspace, and thus lead to new QES potentials.

The characterization of $\cD_2(\cP_n(z))$ was straightforward: an
operator that preserves $\cP_n(z)$ must map polynomials to
polynomials and the same must be true for all
$\mathrm{SL}_2$-related operators.  However, the characterization of
$\cD_2(\cE^{a,b}_n(z))$ is more complicated: the operators in
question must preserve the first order condition \eqref{eq:econd} in
an $\mathrm{SL}_2$-covariant manner. This is addressed by
Propositions \ref{prop:fullE} and \ref{prop:emodT}, which should be
regarded as the  $\mathrm{X}_1$ analogues of Propositions
\ref{prop:Pops} and \ref{prop:liealgT}, respectively.
\begin{proposition}
  \label{prop:fullE}
  A second order $T=p(z) D_{zz} + q(z) D_z + r(z)$ preserves
  $\cE^{a,b}(z)$ if and only if $p(z)$ is a polynomial, and there
  exist polynomials $q_0(z)$ and $r_0(z)$ such that
  \begin{align}
    \label{eq:q0def}
    q(z) &= q_0(z) + p'(z) - \frac{2p(z)}{z-b},\\
    \label{eq:r0def}
    r(z) &= r_0(z)- \frac{2ap(z)}{z-b},\\
    \label{eq:q0cond}
    q_0(b) &= 2 a\, p(b),\\
    \label{eq:r0cond}
    r'_0(b)&= a\,q_0'(b) +a\, p''(b)-a^2\,p'(b).
  \end{align}
\end{proposition}
\noindent
The above Proposition follows directly from the following two Lemmas.
\begin{lemma}
  \label{lem:ceops}
  A second order operator $T$ preserves $\cE^{a,b}(z)$ if and only if it is a
  linear combination of
  \begin{align}
    \label{eq:TE1}
   & D_{zz} + \left( 2a - \frac{2}{z-b}\right) D_z - \frac{2a}{z-b},\\
    \label{eq:TE2}
   & (z-b)D_{zz} +( a(z-b)-1) D_z ,\\
   \label{eq:TE3}
   & (z-b)^k D_{zz},\; k\geq 2,\\
   \label{eq:TE4}
   & (z-b)(D_z+ a),\quad (z-b)^k D_z,\; k\geq 2,\\
   \label{eq:TE5}
   & 1,\quad (z-b)^k,\; k\geq 2.
  \end{align}
\end{lemma}
\begin{proof}
  Showing that the above operators preserve $\cE^{a,b}(z)$ is a a
  straightforward calculation.  We prove the converse.  Since
  $\cE^{a,b}(z)=\cE^{a,0}(z-b)$, no generality will be lost by
  restricting to the case $b=0$; the general case follows by applying
  a translation transformation.  Suppose that $T$ is a second order
  operator that preserves $\cE^{a,0}(z)$, so that $T[az-1], T[z^2],
  T[z^3], T[z^4]$ are all polynomials. By Proposition
  \ref{prop:Tcomponents} and a calculation, it follows that $T$ is a linear
  combination of
  \begin{equation*}
    z^{-1} D_{zz},\quad z^{-2} D_z,\quad z^{-1} D_z,\quad z^{-1},\quad
    z^{-2}
  \end{equation*}
  and an operator with polynomial components.  Hence, $T=T_2+T_1$, where
  $T_2$ is a linear combination of the operators
  \eqref{eq:TE1}-\eqref{eq:TE5}, and $T_1$ is an operator of the
  form
  \begin{equation*}
    T_1 = \frac{p_{-1}}{z} D_{zz} + \left(q_0 + \frac{q_{-1}}{z} +
      \frac{q_{-2}}{z^2}\right) D_z + r_1 z + \frac{r_{-1}}{z} +
    \frac{r_{-2}}{z^2} .
  \end{equation*}
  However, $T_1$ applied to $az-1, z^2, z^3, z^4$ must yield
  polynomials that satisfy \eqref{eq:econd}, which is  only possible if $T_1=0$.
\end{proof}
\begin{lemma}
  \label{lem:fullE}
  A second order operator  satisfies conditions \eqref{eq:q0def} -
  \eqref{eq:r0cond} if and only if it is a linear combination of the
  operators \eqref{eq:TE1}-\eqref{eq:TE5}.
\end{lemma}
\begin{proof}
  One direction of the claim is a straightforward calculation.
  Regarding the converse, let $T$ be an operator that satisfies
  \eqref{eq:q0def} - \eqref{eq:r0cond}. Subtracting the appropriate
  linear combination of \eqref{eq:TE1}-\eqref{eq:TE3} from $T$ yields
  a first order operator that also satisfies \eqref{eq:q0def} -
  \eqref{eq:r0cond}, so it suffices to prove the claim for  a first order $T$ , i.e., we can assume $p(z)=0$. In this case,
  conditions \eqref{eq:q0def} and \eqref{eq:r0def} assert that
  $q(z)$ and
$ r(z)$ are polynomials, while conditions \eqref{eq:q0cond}
  \eqref{eq:r0cond} reduce to
  \begin{equation*}
    q(b) = 0,\quad r'(b) = aq'(b).
  \end{equation*}
  By inspection, the first order $T$ in question must be a linear
  combination of the operators shown in \eqref{eq:TE4}-\eqref{eq:TE5},
  as was to be shown.
\end{proof}
\begin{proposition}
  \label{prop:allofE}
  If a second order $T$ preserves $\cE^{a,b}_n(z),\;n\geq 4$,
  then $T$ preserves the full $\cE^{a,b}(z)$.
\end{proposition}
\begin{proof}
  The proof of Lemma \ref{lem:ceops} applies, mutatis mutandi, to show
  that if $T$ preserves $\cE^{a,b}_n(z)$ for $n\geq 4$, then $T$ is a
  linear combination of the operators
  \eqref{eq:TE1}-\eqref{eq:TE5}, and these operators preserve $\cE^{a,b}(z)$.
\end{proof}
\begin{proposition}
  \label{prop:emodT}
  Let $n\geq 4$ and $a,b$ be given.  Suppose that $b\neq 0$ and set
  $\ha= b(n+ab),\; \hb=-1/b$.  A second order operator $T$ preserves
  $\cE^{a,b}_n(z)$ if and only if $T$ preserves $\cE^{a,b}(z)$ and its
  inversion preserves $\cE^{\ha,\hb}(w)$.
\end{proposition}
\begin{proof}
  Suppose that $T$ preserves $\cE^{a,b}_n(z)$.  By Proposition
  \ref{prop:Ebasis2}, the inversion of $T$, call it $\hT$, preserves
  $\cE^{\ha,\hb}_n(w)$.  The desired conclusion now follows by
  Proposition \ref{prop:allofE}.
To prove the converse, let $f(z)\in\cE^{a,b}_n(z)$ be given and
  set
  \begin{equation*}
    g(z)=T[f(z)],\quad \hf(w) = w^n f(-1/w)\in\cE^{\ha,\hb}_n(w),\quad
    \hg(w)=\hT[\hf(w)].
  \end{equation*}
  By assumption, $g(z)\in\cE^{a,b}(z)$ while
  $\hg(w)\in\cE^{\ha,\hb}(w)$, and by covariance,
  \begin{equation*}
    \hg(w) = w^n g(-1/w).
  \end{equation*}
  This implies that $\deg g\leq n$, and therefore, $T$ preserves
  $\cE^{a,b}_n(z)$.
\end{proof}
\begin{proposition}
  \label{prop:Eop}
  If a second order operator $T$ preserves  $\cE_n^{a,b}(z)$, then
  necessarily $\deg p(z)\leq 4$.  Furthermore, a basis of
  $\cD_2(\cE_n^{a,b}(z))$ is given by the following seven operators:
  \begin{align}
    \label{eq:jops1}
    J_1 &= (z-b)^4D_{zz} - 2(n-1)(z-b)^3D_z +n(n-1)(z-b)^2, \\
    \label{eq:jops2}
    J_2   &= (z-b)^3 D_{zz} -(n-1) (z-b)^2D_z\;,\\
    \label{eq:jops3}
    J_3 &= ( z-b)^2D_{zz} \;,\\
    \label{eq:jops4}
    J_4  & = ( z-b)D_{zz} + \lp a(z-b)-1\rp D_z \;,\\
    \label{eq:jops5}
    J_5 &=D_{zz}+2\lp a-\frac{1}{z-b}\rp \,D_z - \frac{2a}{
      z-b}\;,\\
    \label{eq:jops6}
    J_6 &= (z-b)\lp z-b-\frac{n}{a}\rp D_z -n (z-b) ,\\
    \label{eq:jops7}
    J_7 &= 1.
  \end{align}
\end{proposition}
\begin{proof}
  There are several ways to obtain the above basis.  The proof below
  is based on $\mathrm{SL}_2$ covariance, c.f., the proof of Proposition
  \ref{prop:labasis}.  See \cite{GKM3} for a different approach.

  Without loss of generality, we assume that $b\neq 0$. The above
  operators satisfy conditions \eqref{eq:q0def}-\eqref{eq:r0cond}, and
  hence preserve $\cE^{a,b}(z)$. Using
  \eqref{eq:inversion-p}-\eqref{eq:inversion-r}, a calculation
  shows that the inversions of these operators obey
  \eqref{eq:q0def}-\eqref{eq:r0cond} and hence preserve
  $\cE^{\ha,\hb}(w)$, where $\ha, \hb$ are given by \eqref{eq:Ebasis2}.
  Therefore, by Proposition \ref{prop:emodT}, the
  above 7 operators preserve $\cE^{a,b}_n(z)$.

  Let us now prove that the above operators span
  $\cD_2(\cE^{a,b}_n(z))$.  Let $T=p(z)D_{zz}+q(z) D_z + r(z)$ be a
  second order operator that preserves $\cE^{a,b}_n(z)$ and let
  $\hT=\hp(w)D_{ww}+\hq(w) D_w + \hr(w)$ denote the
  inversion of $T$.  The latter preserves $\cE^{\ha,\hb}_n(w)$ by Proposition
  \ref{prop:Ebasis2}.  By
  Propositions \ref{prop:fullE} and \ref{prop:allofE}, both $p(z)$ and
  $\hp(w)$ are polynomials, whence by \eqref{eq:inversion-p},
  necessarily, $\deg p \leq 4$.  Hence, by subtracting an
  appropriate linear combination of the operators $J_1, J_2, J_3, J_4,
  J_5$ we obtain a first order operator $T_1=q_1(z)D_z + r_1(z)$ that
  preserves $\cE^{a,b}_n(z)$.  Let $\hT_1=\hq_1(w) D_w+ \hr_1(w)$
  denote the inversion of $T_1$, where by \eqref{eq:inversion-q}
  \eqref{eq:inversion-r} we have
  \begin{equation*}
    \hq_1(w) = w^2 q_1(-1/w),\quad \hr_1(w) = r(-1/w)-nwq(-1/w).
  \end{equation*}
  By Propositions \ref{prop:fullE} and \ref{prop:allofE}, $q_1(z),
  \hq_1(w), r_1(z)$ and $\hr_1(w)$ are polynomials; hence $\deg q_1
  \leq 2$ while $\deg r_1 \leq 1$.  By Lemma \ref{lem:fullE},
  $T_1$ must be a linear combination of $(z-b)(D_z+a)$, $(z-b)^2 D_z$, and
  a constant.  The requirement that $\hr_1(w)$ be a polynomial
  forces $T_1$ to be a multiple of $J_6$ plus a constant.
\end{proof}

We shall refer to the potential invariant $v(z)$ of an
$\mathrm{X}_1$ operator as an {\em $\mathrm{X}_1$ potential}. We can
now state the main theoretical result of the paper: the
characterization of $\mathrm{X}_1$ potentials. It turns out that it
is easier to describe  $\mathrm{X}_1$ potentials than $\mathrm{X}_1$
operators as evinced by the following\footnote{ It is instructive to
compare the characterization of $\mathrm{X}_1$ potentials given in
Theorem \ref{thm:e1111}  with the corresponding characterization of
Lie algebraic potentials given in Theorem \ref{thm:v1111}.}
\begin{theorem}
  \label{thm:e1111}
  Let $p(z)$ be a fourth degree polynomial with distinct roots $z_i$, and
  let $b$ be a constant such $p(b)\neq 0$.  Let $k_1,k_2,k_3$ and $k_4$ be
  constants such that
  \begin{align}
    \label{eq:esumk}
    k_1+k_2+k_3+k_4&=-n,\\
    \label{eq:esumk1}
    \sum_{i=1}^4 \frac{k_i^2 p'(z_i)}{(b-z_i)^2} &= 0.
  \end{align}
  Then, the rational function
  \begin{equation}
    \label{eq:e1111}
    v(z) =-\frac{2\,p(b)}{(z-b)^2}-\frac{p\,{}'\!(b)}{z-b}
    -\sum_{i=1}^4 \lp k_i{}^2-\tfrac{1}{16}\rp
    \,\frac{p'(z_i)}{z-z_i}+\lambda,
  \end{equation}
  is an $\mathrm{X}_1$ potential.
  The corresponding $\mathrm{X}_1$ operator
  preserves $\cE^{a,b}_n(z)$, where
  \begin{equation}
    \label{eq:akrel}
    a = \sum_{i=1}^4 \frac{k_i}{b-z_i}.
  \end{equation}
  Conversely, let $T=p(z) D_{zz}+q(z) D_z+r(z)$ be a second-order
  operator such that $p(z)$ is a fourth degree polynomial with distinct
  roots $z_i$. Let $v(z)$ denote the corresponding potential, as given
  by \eqref{eq:vform}, and let
  \begin{equation}
    \label{eq:kpqrel}
    k_i = \frac{q(z_i)}{2 p'(z_i)} - \frac{1}{2}.
  \end{equation}
  If $T$ is an $\mathrm{X}_1$ operator preserving
  $\cE^{a,b}_n(z)$, then $p(b)\neq 0$, and equations
  \eqref{eq:esumk}, \eqref{eq:esumk1}, \eqref{eq:e1111},
  \eqref{eq:akrel} all hold.
\end{theorem}
\noindent The proof of the theorem proceeds via a number of
Lemmas. The following Lemma identifies those operators that are
both Lie-algebraic and preserve $\cE^{a,b}_n(z)$.  Above, we
adopted the convention that an operator cannot be both
$\mathrm{X}_1$ and Lie-algebraic.  Thus, some Lie-algebraic
operators preserve an $\mathrm{X}_1$ subspace, but by definition,
an $\mathrm{X}_1$ operator must not preserve $\cP_n(z)$.
\begin{lemma}
  \label{lem:pbnotzero}
  Let $T=p(z) D_{zz}+q(z) D_z+r(z)$ be a second order operator such that
  $p(z)$ is a fourth degree polynomial with distinct roots $z_i$.  If $T$
  preserves $\cE^{a,b}(z)$, then $p(b)=0$ if and only if $T$ also
  preserves $\cP(z)$. Similarly, if $T$ preserves $\cE^{a,b}_n(z)$,
  then $p(b)=0$ if and only if $T$ also preserves $\cP_n(z)$.
\end{lemma}
\begin{proof}
  Suppose that $T$ preserves $\cE^{a,b}(z)$. If $T$ also preserves
  $\cP(z)$, then by Proposition \ref{prop:Pops}, $q(z)$ is a
  polynomial.  Hence, by equation \eqref{eq:q0def} of Proposition
  \ref{prop:fullE}, we must have $p(b)=0$.  Conversely, suppose that
  $p(b)=0$.  By \eqref{eq:q0def} and \eqref{eq:r0def} both $q(z)$ and
  $r(z)$ are polynomials. Hence, $T$ preserves $\cP(z)$, by
  Proposition \ref{prop:allofP}.

  Next, suppose that $T$ preserves the finite-dimensional
  $\cE^{a,b}_n(z)$, and let $\hT$ denote the inversion of $T$. If $T$
  preserves $\cP_n(z)$, then the above argument shows that
  $p(b)=0$.  Conversely, suppose that $p(b)=0$.  A repeat of the above
  argument shows that $T$ preserves $\cP(z)$.  A calculation based on
  \eqref{eq:inversion-p}-\eqref{eq:inversion-q} shows that
  $\hp(\hb)=0$ where $\hb=-1/b$.  By Proposition \ref{prop:Ebasis2},
  $\hT$ preserves $\cE^{\ha,\hb}_n(z)$, and therefore the transformed components $\hq(w)$ and $\hr(w)$ are
  polynomials in $w$.  Proposition \ref{prop:liealgT} then implies
  that $T$ preserves $\cP_n(z)$.
\end{proof}

\begin{lemma}
  \label{lem:esumk1}
  Let $p(z)$ be a fourth degree polynomial with distinct roots $z_i$ and
  $q_0(z)$ be a polynomial. Let $a$ and $b$ be constants such that equation
  \eqref{eq:q0cond} holds and  $p(b)\neq 0$. Let $T=p(z)
  D_{zz}+q(z) D_z+r(z)$ be a second order operator where $q(z)$ is given
  by \eqref{eq:q0def} and where $r(z)$ is a rational function. Let
  $k_i$ be defined by \eqref{eq:kpqrel} and let $\Delta(z)$ denote the
  difference of the right-hand side expressions in \eqref{eq:vform}
  and \eqref{eq:e1111}.  Then, $T$ preserves $\cE^{a,b}(z)$ if and
  only if $\Delta(z)$ is a polynomial such that
  \begin{equation}
    \label{eq:deltab}
    \Delta'(b) = - \sum_{i=1}^4 \frac{k_i^2 p'(z_i)}{(b-z_i)^2}.
  \end{equation}
\end{lemma}
\begin{proof}
  We begin by listing some consequences of the above assumptions.
  Define $r_0(z)$ so that equation \eqref{eq:r0def} holds.
  Substituting \eqref{eq:q0def} \eqref{eq:r0def} into the given
  expression for $\Delta(z)$ gives
  \begin{align*}
    \Delta(z) &= r_0(z)- \frac{2ap(z)}{z-b} -\frac{1}{2} \left(q_0{}'(z) +
      p''(z) -
      \frac{2p'(z)}{z-b}+\frac{2p(z)}{(z-b)^2}\right)\\
    &\qquad +\frac{1}{4} p''(z) -\frac{1}{4}\frac{\lp q_0(z)(z-b) -
      2p(z) \rp^2}{p(z)(z-b)^2}+
    \frac{1}{16}\frac{p'(z)^2}{ p(z)}\\
    &\qquad+\frac{2\,p(b)}{(z-b)^2}+\frac{p\,{}'\!(b)}{z-b}
    +\sum_{i=1}^4 \lp k_i{}^2-\tfrac{1}{16}\rp
    \,\frac{p'(z_i)}{z-z_i}-\lambda.
  \end{align*}
  Rearranging, and using \eqref{eq:q0cond}, the above expression can be written as
  \begin{align}
    \label{eq:esumk1a}
    \Delta(z)&= r_0(z)-
    \frac{1}{2}q_0'(z)-\frac{1}{4}p''(z) -
    2a\left(\frac{p(z)-p(b)}{z-b}\right) \\
    \nonumber &\qquad  + \frac{q_0(z)-q_0(b)}{z-b}+
    \frac{p'(z)-p'(b) }{z-b}- 2\left( \frac{p(z)
        -p(b) - (z-b)\,p'(b)}{(z-b)^2}\right)  \\ \nonumber &\qquad
    + \frac{1}{16}\frac{{p'(z)}^2}{p(z)}
    - \frac{1}{4}\frac{q_0(z)^2}{p(z)}
    +\sum_{i=1}^4 \lp k_i{}^2-\tfrac{1}{16}\rp
    \,\frac{p'(z_i)}{z-z_i}-\lambda.
  \end{align}
  Applying the identity
  \begin{equation}
    \label{eq:pident}
    D_z\lp \frac{{p'(z)}^2}{p(z)} \rp\Bigg|_{z=b}= \frac{4}{3}\,
    p^{(3)}(b)+\sum_{i=1}^4 \frac{p'(z_i)}{(b-z_i)^2},
  \end{equation}
  using \eqref{eq:q0cond}, and canceling
  gives
  \begin{equation}
    \label{eq:esumk1b}
    \Delta'(b) = \Big( r_0'(b) - a\, p''(b)+ a^2 p'(b) -a
      q_0'(b)\Big)- \sum_{i=1}^4
    \frac{k_i{}^2 p'(z_i)}{(b-z_i)^2}.
  \end{equation}
  Note that \eqref{eq:r0cond} describes the vanishing of the bracketed
  term in the right hand side of the above equation.  As well,
  \eqref{eq:kpqrel} and \eqref{eq:q0def} give
  \begin{equation}
    \label{eq:kpq0rel}
    k_i = \frac{q_0(z_i)}{2 p(z_i)}.
  \end{equation}
  We also note that, by equation \eqref{eq:esumk1a}, $\Delta(z)$ has vanishing
  residues at $z=z_i$.

  Now, suppose that $T$ preserves $\cE^{a,b}(z)$.  By Proposition
  \ref{prop:fullE}, $r_0(z)$ is a polynomial that satisfies equation
  \eqref{eq:r0cond}. Therefore, by equations \eqref{eq:esumk1a}
  \eqref{eq:esumk1b} \eqref{eq:kpq0rel}, $\Delta(z)$ is a polynomial such
  that condition \eqref{eq:deltab} holds.  Conversely, suppose that
  $\Delta(z)$ is a polynomial such that \eqref{eq:deltab} holds.
  Since $\Delta(z)$ has no poles, equations \eqref{eq:esumk1a}
  \eqref{eq:kpq0rel} imply that $r_0(z)$ is a polynomial.  Finally,
  equations \eqref{eq:deltab} and \eqref{eq:esumk1b} imply that
  \eqref{eq:r0cond} holds. Therefore, by Proposition \ref{prop:fullE},
  $T$ preserves $\cE^{a,b}(z)$.
\end{proof}
\noindent
The following Lemmas establish additional transformation rules for
the inversion $z=-1/w$.
\begin{lemma}
  \label{lem:e1111inv}
  Let $p(z)$ be a fourth degree polynomial with distinct roots $z_i\neq
  0$; let $b$ be a constant such that $p(b)\neq 0$; let $v(z)$ be
  defined by \eqref{eq:e1111}; set $z=-1/w$, $\hb=-1/b$, $w_i =
  -1/z_i$, $\hv(w)=v(-1/w)$.  Then,
  \begin{align}
    \label{eq:e1111inv-3}
    \hv(w) &= -\frac{2\,\hp(\hb)}{(w-\hb)^2}
    -\frac{\hp'(\hb)}{w-\hb} -\sum_{i=1}^4 \lp
    k_i{}^2-\tfrac{1}{16}\rp
    \,\frac{\hp'(w_i)}{w-w_i}+\lambda_0,\intertext{where}
    \nonumber
    \lambda_0 &= \frac{2\hp(\hb)}{\hb^2} -
    \frac{\hp'(\hb)}{\hb} - \sum_{i=1}^4  \lp
    k_i{}^2-\tfrac{1}{16}\rp \frac{\hp'(w_i)}{w_i}.
  \end{align}
  As well,
  \begin{equation}
    \label{eq:e1111inv-4}
    \sum_{i=1}^4 \frac{k_i^2\, p'(z_i)}{(b-z_i)^2} = \hb^2
    \sum_{i=1}^4 \frac{k_i^2\, \hp'(w_i)}{(\hb-w_i)^2}.
  \end{equation}
\end{lemma}
\begin{proof}
  Applying the above substitutions and \eqref{eq:inversion-p'}  yields
  \begin{align*}
    -\frac{2\,p(b)}{(z-b)^2}-\frac{p'(b)}{z-b} &=
    -\frac{2\,\hp(\hb)}{(w-\hb)^2}-\frac{\hp'(\hb)}{w-\hb} +
    \frac{2\hp(\hb)}{\hb^2} - \frac{\hp'(\hb)}{\hb},\\
    \frac{p'(z_i)}{z-z_i} &= \frac{\hp'(w_i)}{w-w_i} +
    \frac{\hp'(w_i)}{w_i},
  \end{align*}
from which equations \eqref{eq:e1111inv-3} \eqref{eq:e1111inv-4}
follow immediately.
\end{proof}
\begin{lemma}
  \label{lem:esumk}
  Let $p(z)$ be a fourth degree polynomial with distinct roots $z_i\neq
  0$; let $\hp(w)$ be the polynomial defined by \eqref{eq:inversion-p}
  and let $q(z)$ and $\hq(w)$ be rational functions related by
  \eqref{eq:inversion-q}; let $q_0(z)$ be defined by equation
  \eqref{eq:q0def}, and let $\hq_0(w)$ be defined by the analogous
  equation:
  \begin{equation}
    \label{eq:q0wdef}
      \hq(w) = \hq_0(w) + \hp'(w) - \frac{2 \hp(w)}{w-\hb},
  \end{equation}
  where $w=-1/z, \hb=-1/b$.  Then, \eqref{eq:kpqrel} holds and
  $q_0(z), \hq_0(w)$ are both polynomials if and only if
  \eqref{eq:esumk} holds and if $q_0(z)$ obeys
  \begin{equation}
    \label{eq:qzpzemod}
    \frac{q_0(z)}{2p(z)}=\sum_{i=1}^4 \frac{k_i}{z-z_i}.
  \end{equation}
\end{lemma}
\begin{proof}
  Using equations \eqref{eq:inversion-p} \eqref{eq:inversion-q}
  \eqref{eq:inversion-p'} \eqref{eq:q0def}, a calculation shows that
  \begin{equation}
    \label{eq:q0zwrel}
    \frac{\hq_0(w)}{\hp(w)} = \frac{1}{w^2} \,
    \frac{q_0(-1/w)}{p(-1/w)}
    - \frac{2n}{w}\,.
  \end{equation}
  Suppose now  that \eqref{eq:kpqrel} holds and that $q_0(z)$ and $\hq_0(w)$
  are both polynomials.  By equation \eqref{eq:q0zwrel}, this is only
  possible if $\deg q_0\leq 3$.  Equation \eqref{eq:kpqrel} gives
  equation \eqref{eq:kpq0rel}, which gives
  equation \eqref{eq:qzpzemod}.

  Conversely, suppose that equations \eqref{eq:esumk} and
  \eqref{eq:qzpzemod} hold.  Taking the residues of the latter at
  $z=z_i$ gives equation \eqref{eq:kpq0rel}, which gives equation
  \eqref{eq:kpqrel}.  Using equations \eqref{eq:esumk}
  \eqref{eq:qzpzemod} \eqref{eq:q0zwrel} and making the substitutions
  $z=-1/w, z_i = -1/w_i$ gives
  \begin{align}
    \nonumber
    \frac{\hq_0(w)}{2\hp(w)} &= \sum_{i=1}^4 \frac{k_i}{w-w_i}
    -\left(n+\sum_{i=1}^4 k_i\right)
    w^{-1} \\ \label{eq:qwpwemod}
    &= \sum_{i=1}^4 \frac{k_i}{w-w_i}.
  \end{align}
  Therefore, $\hq_0(w)$ is also a polynomial.
\end{proof}

\begin{proof}[Proof of theorem \ref{thm:e1111}]
  Suppose that $p(z)$ and $b$ satisfy the stated assumptions, that $k_1,
  k_2, k_3, k_4$ are constants satisfying equations \eqref{eq:esumk}
  and \eqref{eq:esumk1}, and that $v(z)$ has the form shown in
  \eqref{eq:e1111}.  We assume, without loss of generality, that
  $p(0)\neq 0$.  Define the polynomial $q_0(z)$ by \eqref{eq:qzpzemod}
  and set $q(z)$ so that \eqref{eq:q0def} holds. Define the constant
  $a$ by equation \eqref{eq:akrel}; evaluating \eqref{eq:qzpzemod} at
  $z=b$ then shows that equation \eqref{eq:q0cond} holds.  Consider
  the operator $T=p(z) D_{zz} + q(z) D_z + r(z)$, where $r(z)$ is
  defined so that \eqref{eq:vform} holds. By construction,
  $\Delta(z)=0$, where $\Delta(z)$ denotes the difference between the
  right-hand sides of \eqref{eq:vform} and \eqref{eq:e1111}.  Hence,
  by Lemma \ref{lem:esumk1}, $T$ preserves $\cE^{a,b}(z)$.  Let
  $\hT=\hp(w) D_{ww} + \hq(w) D_w + \hr(w)$ be the inversion of $T$,
  and let $\hv(w)=v(-1/w)$ denote the corresponding potential.  By
  Lemma \ref{lem:e1111inv}, $\hv(w)$ has the form shown in
  \eqref{eq:e1111} with $w, \hp(w), w_i = -1/z_i,$ and $\hb=-1/b$ replacing
  $z, p(z), z_i,$ and $b$ , respectively.  Define $q_0(w)$ so that equation
  \eqref{eq:q0wdef} holds. By Lemma \ref{lem:esumk}, $q_0(w)$ is a
  polynomial having the form shown in \eqref{eq:qwpwemod}.  Hence,
  mutatis mutandi, we can apply the above argument to conclude that
  $\hT$ preserves $\cE^{\ha,\hb}(w)$, where
  \begin{equation}
    \label{eq:hakrel}
    \ha = \sum_{i=1}^4 \frac{k_i}{\hb-w_i} = b\sum_{i=1}^4 \left(
    \frac{k_i}{b-z_i} - k_i\right)= b(n+ab).
  \end{equation}
  Therefore, by Proposition \ref{prop:emodT}, the operator $T$
  preserves $\cE_n^{a,b}(z)$.  By Lemma \ref{lem:pbnotzero}, $T$ does
  not preserve $\cP_n(z)$, and is, therefore, an $\mathrm{X}_1$ operator.

  Conversely, suppose that $T=p(z) D_{zz} + q_z D_z + r(z)$ is an
  operator that preserves $\cE^{a,b}_n(z)$, but does not preserve
  $\cP_n(z)$.  By Lemma \ref{lem:pbnotzero}, $p(b)\neq 0$. By
  Proposition \ref{prop:allofE}, $T$ preserves $\cE^{a,b}(z)$.  Hence,
  Proposition \ref{prop:fullE} applies, and we can define polynomials
  $q_0(z), r_0(z)$ so that equations
  \eqref{eq:q0def}-\eqref{eq:r0cond} hold.  Let $k_i$ be defined by
  equation \eqref{eq:kpqrel}, and let $\Delta(z)$ denote the
  difference of the expressions shown in \eqref{eq:vform} and
  \eqref{eq:e1111}.  By Lemma \eqref{eq:esumk1}, $\Delta(z)$ is a
  polynomial such that equation \eqref{eq:deltab} holds.  Let
  $\hT=\hp(w) D_{ww}+\hq(w) D_w + \hr(w)$ denote the inversion of $T$.
  By covariance and by Proposition \ref{prop:Ebasis2}, $\hT$ preserves
  $\cE^{\ha,\hb}_n(w)$ where
  \begin{equation*}
    w=-1/z, \quad \ha= n(b+ab),\quad \hb = -1/b.
  \end{equation*}
  Applying the above argument, mutatis mutandi, to $\hT$ we conclude
  that $\hDelta(w)$ is a polynomial such that equation
  \eqref{eq:deltab} holds with $p(z), b, z_i$ replaced by $\hp(w),
  \hb, w_i=-1/z_i$, respectively.  By Lemma \ref{lem:e1111inv},
  $\hDelta(w)$ and $\Delta(-1/w)$ differ by a constant.  Since both
  $\hDelta(w)$ and $\Delta(z)$ are polynomials, a fortiori, they must
  both be constants. Hence, the potential $v(z)$ has the form shown in
  \eqref{eq:e1111}.  Furthermore, since $\Delta'(b)=0$, equation
  \eqref{eq:deltab} implies \eqref{eq:esumk1}.  Finally, since both
  $q_0(z)$ and $\hq_0(w)$ are polynomials and since \eqref{eq:kpqrel}
  holds, Lemma \ref{lem:esumk} gives equations \eqref{eq:esumk} and
  \eqref{eq:qzpzemod}.  Evaluating the latter at $z=b$, and using
  \eqref{eq:q0cond} gives equation \eqref{eq:akrel}.
\end{proof}
\section{Classification of $\mathrm{X}_1$
potentials.}\label{sec:classification}

The characterization of $\mathrm{X}_1$ potentials achieved by
Theorem \ref{thm:e1111} allows to perform a complete classification
of these potentials, by using a similar approach to the
classification of Lie-algebraic potentials. Since the equivalence
problem admits $\mathrm{SL}_2$ covariance and the leading order
component of the operator $T$ is a quartic, the classification
proceeds by considering the 6 possible root configurations of a
quartic polynomial.  As above, we name the classes type I, II, D, Z,
III, and N.

To keep things manageable we focus on non-singular potentials
whose domain, in the physical variable $x$, is the entire real
line. The type Z change to physical coordinate is $z=\sinh x$, so
that both the physical and the algebraic domain are the entire
real line. However, an $\mathrm{X}_1$ potential possesses a term
of the form $1/(z-b)$, where $b\in\Rset$, and therefore
$\mathrm{X}_1$ potentials of type Z are necessarily singular.  A
similar line of reasoning holds for  type N potentials.  That
leaves potentials of type I, II, D, and III, which can be obtained
by specializing the values of the roots  $\rho_i$ in Theorem
\ref{thm:e1111}.  In the case of multiple roots, the potential
form and the constraints are obtained as the appropriate limit of
the generic form \eqref{eq:e1111}.

Some of the resulting potentials also possess multiple algebraic
sectors, the analysis being essentially the same as for the
Lie-algebraic case.
\subsection{Type I potentials.}
The $\mathrm{X}_1$ potentials of type I are deformations of the
generalized Lam\'e potentials.  We take
\begin{equation*}
  p(z) = 4z(z-1)(mz-m+1),\quad m\in(0,1),
\end{equation*}
and denote the roots of $p(z)$ by
\begin{equation*}
z_1=1-\frac{1}{m},\quad  z_2=0, \quad z_3=1,\quad z_4=\infty.
\end{equation*}
The physical coordinate $x$ is related to the algebraic coordinate $z$
by $z=\cn(x| m)^2$, where $\cn(x| m)$ is the Jacobi elliptic function
of modulus $m$.  In this way, the physical domain, $-\infty<x<\infty$,
maps onto the interval $0\leq z\leq 1$.  We now take the potential
form shown in Theorem \ref{thm:e1111}, perform an inversion $z=-1/w$
and to ensure that the potential is non-singular demand that $k_3=k_4 =
\pm 1/4$, and that $b\not\in [0,1]$. The resulting potential has the
following form:
\begin{align}
  \label{eq:emodlamepot}
  u(x) =&\ m\,c_1(c_1+1)\, \sn(x| m)^2 + m\, c_2(c_2+1)\,
  \frac{\cn(x| m)^2}{\dn(x| m)^2}+\\ \nonumber & -
  \frac{p'(b)}{\cn(x| m)^2-b} - \frac{2\, p(b)}{(\cn(x|
    m)^2-b)^2},\qquad c_i=\frac{1}{2}-2k_i.
\end{align}
Restating \eqref{eq:esumk}
\eqref{eq:esumk1}, the constraints on  $k_1, k_2$ are given by
\begin{gather}
  \label{eq:k1k2cond1}
  k_1+k_2=-n-(k_3+k_4),\quad k_3+k_4\in\{-\tfrac{1}{2},0,\tfrac{1}{2}\},\\
  \label{eq:k1k2cond2}
  k_1{}^2+\frac{m-1}{(bm-m+1)^2}\,k_2{}^2=
  \frac{m (b-1)^2+2 b-1}{16 \,(b-1)^2\, b^2\, m}
\end{gather}
Let $K(m), K'(m)=i K(1-m)$ denote the real and imaginary
quarter-periods of $\cn(x|m)$. For $b>1$, the equation
$\cn(x|m)2=\sqrt{b}$ has a unique, up to addition of an integer
multiple of $2K'$, solution of the form $x=i\rho, \rho\in
\mathbb{R}$. A calculation shows that the residue of $u(x)$ vanishes
at $x=\pm i \rho$.  Hence, up to a constant term, the potential can
be expressed as
\begin{align}
  u(x) &= m\{ c_1(c_1+1) \sn(x|m)^2 + c_2 (c_2 +1) \sn(x+K(m) | m)^2 \\
  \nonumber   &\quad +
  2 \sn(x+i \rho|m)^2 +
  2 \sn(x-i \rho|m)^2\, \}.
\end{align}
For integer values of $c_1, c_2$ the resulting expression is
a finite-gap potential (also known as a Picard potential
\cite{GW98}) that was first described in reference \cite{TV97}. It
is the simplest known example of a finite gap potential that falls
outside the elliptic Treibich-Verdier class (the type I Lie
algebraic potentials).  For more on this see reference
\cite{Sm06}.

The algebraic form of the operator is
\begin{equation}
  \label{eq:k1k2T}
  T = \frac{p^{(3)}(b)}{6} J_2  +\frac{p^{(2)}(b)}{2} J_3
  +p'(b) J_4 +p(b) J_5 + \lp k_4+4 mn\rp J_6,
\end{equation}
and it preserves the $\mathrm{X}_1$ subspace $\cE^{a,b}_n(z)$,
where
\begin{equation}
  \label{eq:k1k2aformula}
  a=\frac{mk_2}{mb-m+1}+ \frac{k_3}{b}+\frac{k_4}{b-1}.
\end{equation}
The gauge factor is
\begin{equation}
  \label{eq:k1k2gaugefactor}
  \mu(x) = (\cn(x| m)^2-b)^{-1} \dn(x| m)^{c_2} \cn(x| m)^{c_3}
  \sn(x| m)^{c_4},
\end{equation}
where $c_3, c_4\in\{0,1\}$ according to the signs of $k_3, k_4$.

\subsection{Type II potentials.}
The double root is at infinity.  The other two roots are at $z_3=0$
and $z_4=1$. Thus,
\begin{equation*}
  p(z) = 4z(1-z),\quad   z = -\sinh^2(x)
\end{equation*}
The non-singularity condition is $k_3=\pm 1/4$ and
$b\not\in[-1,0]$. The potential form and the gauge factor are
given by
\begin{gather}
  \begin{aligned}
    u(x) = 4& \ell_1{}^2 \,\sinh^4 x + 4
    \ell_1 (\ell_1+2\ell_2) \,\sinh^2 x \\
    &+c_4(c_4-1)\sech^2 x +
    \frac{8\,b\,(b-1)}{( \sinh^2 x+b)^2} - \frac{4(2\,b-1)}{\sinh^2 x +
      b},
  \end{aligned}\\
  \mu(x) = \exp(-\ell_1 \sinh^2 x) (\sinh^2 x+b)^{-1} (\cosh x)^{c_4}
  (\sinh x)^{c_3},
\end{gather}
where, as before, $c_i= 2k_i+1/2$ and $c_3\in\{0,1\}$.  The potential
parameters are constrained as follows:
\begin{gather}
  k_4+\ell_2=-n-k_3,\quad k_3 =\pm \frac{1}{4},\\
  \ell_1\left((2\,b-1)\ell_1-2\ell_2\right)-\frac{k_4{}^2}{(b-1)^2} -
  \frac{1}{16\, b^2}=0
\end{gather}
The algebraic operator
\begin{equation}
  T= -4J_3-4(2\,b-1)J_4-4\,b(b-1)J_5-8\ell_1 J_6
\end{equation}
preserves $\cE^{a,b}_n(z)$, where
\begin{equation}
  a= \frac{k_3}{b}+\frac{k_4}{b-1}+\ell_1.
\end{equation}

\subsection{Type D potentials}
One  double root is at infinity; the other double root is at zero.  Thus,
\begin{gather}
  p(z) = -z^2, \quad   z = \exp(x).
\end{gather}
The potential  and gauge factor are
\begin{align}
    u(x) &= (\ell_1 e^x - 2\ell_2)^2 + (\ell_3 e^{-x} + 2\ell_4)^2 +
    \frac{2b}{e^x -b}+\frac{2b^2}{(e^x-b)^2},\\
  \mu(x) &= (e^x-b)^{-1} \exp(\ell_1 e^x - \ell_3\,
  e^{-x}+(2\ell_4+\tfrac{1}{2})\,x),
\end{align}
where $b<0$ to avoid singularities.
The potential parameters are constrained as follows:
\begin{gather}
  2\ell_2+2\ell_4=-n,\\
  2\ell_3(\ell_3+2b\,\ell_4)-b^3\ell_1(2b\,\ell_1-4\ell_2)=0.
\end{gather}
The algebraic form of the operator is
\begin{equation}
  T= -J_3-2bJ_4-b^2 J_5-2 \ell_1 \, J_6,
\end{equation}
and it preserves the $\mathrm{X}_1$ subspace $\cE^{a,b}_n(z)$,
where
\begin{equation}
  a= \ell_1+\frac{\ell_3}{b^2}+\frac{2\ell_4}{b}.
\end{equation}

\subsection{Type III potentials}
There is a triple root at infinity and a single root  at $z_4=0$.
Thus,
\begin{gather}
  p(z) = -4z, \quad z = x^2.
\end{gather}
The non-singularity condition is $k_4=\pm1/4$. The potential  and
gauge factor are
\begin{align}
  u(x) &= \ell_3{}^2\, x^6-2\ell_2\ell_3\, x^4+4(\ell_2{}^2+\ell_1
  \ell_3)\, x^2+\frac{8b}{(x^2-b)^2}  +\frac{4}{x^2-b}\\
  \mu(x) &=(x^2-b)^{-1} \exp\left(-\frac{\ell_3}{2} x^4 + \ell_2
    x^2\right) x^{c_4},
\end{align}
where, as before, $b<0$, $c_4= 2k_4+1/2$ and $c_4\in\{0,1\}$.  The potential
parameters are constrained as follows:
\begin{gather}
  \ell_1 = -n-k_4,\quad k_4 =\pm \frac{1}{4},\\
  \frac{3}{4}\,b^2\, \ell_3{}^2 -2b\, \ell_2\ell_3+\ell_1\ell_3+\ell_2{}^2
   = \frac{k_4{}^2}{b^2}.
\end{gather}
The algebraic form of the operator is
\begin{equation}
  T= -4J_4-4bJ_5+8 \ell_3 J_6,
\end{equation}
and it preserves the $\mathrm{X}_1$ subspace $\cE^{a,b}_n(z)$,
where
\begin{equation}
  a= \ell_2-b\ell_3+\frac{k_4}{b}.
\end{equation}

\section{An example of an $\mathrm X_1$ elliptic potential.}\label{sec:example}
\begin{figure}[h]
\begin{center}
\begin{tabular}{cc}
\begin{tabular}{c}
\psfig{figure=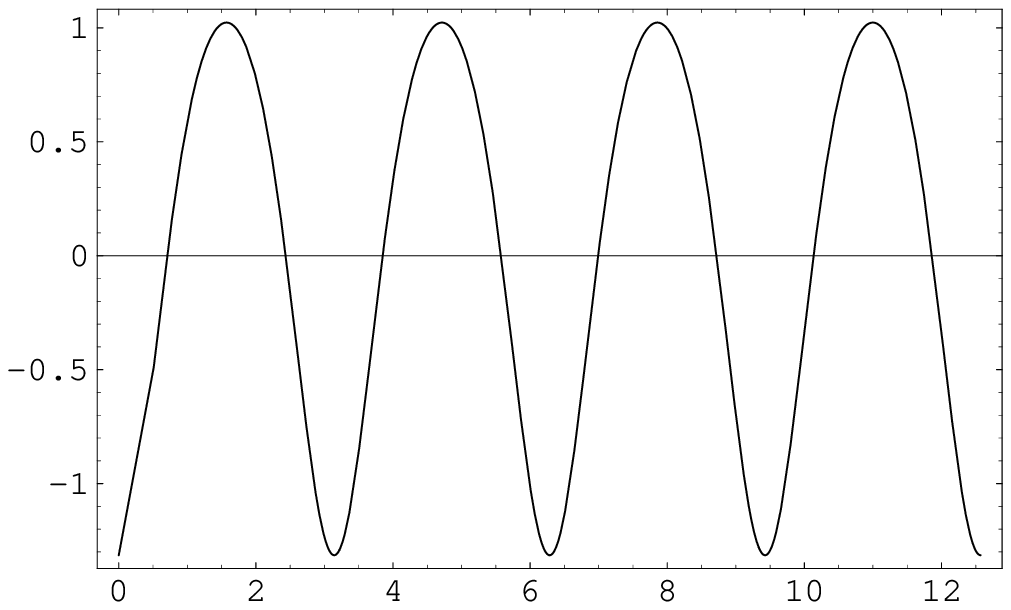,width=2.2in}\\{ \scriptsize{Fig. 1a$\qquad
m=0.01$}}
\end{tabular} &
\begin{tabular}{c}
\psfig{figure=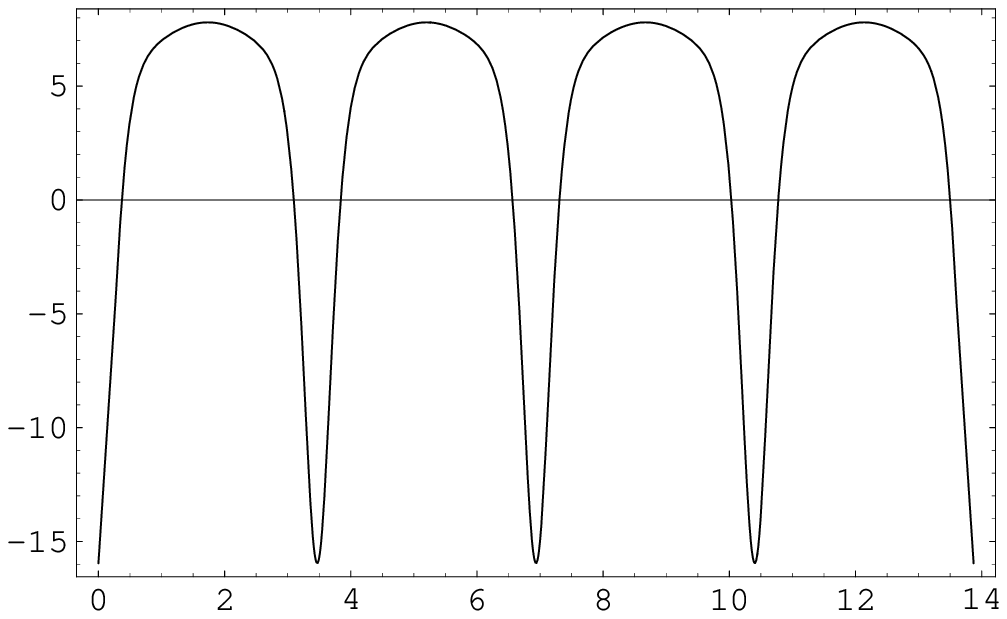,width=2.2in}\\ {\scriptsize{Fig. 1b$\qquad
m=0.33$}}
\end{tabular}\\\\
\begin{tabular}{c}
\psfig{figure=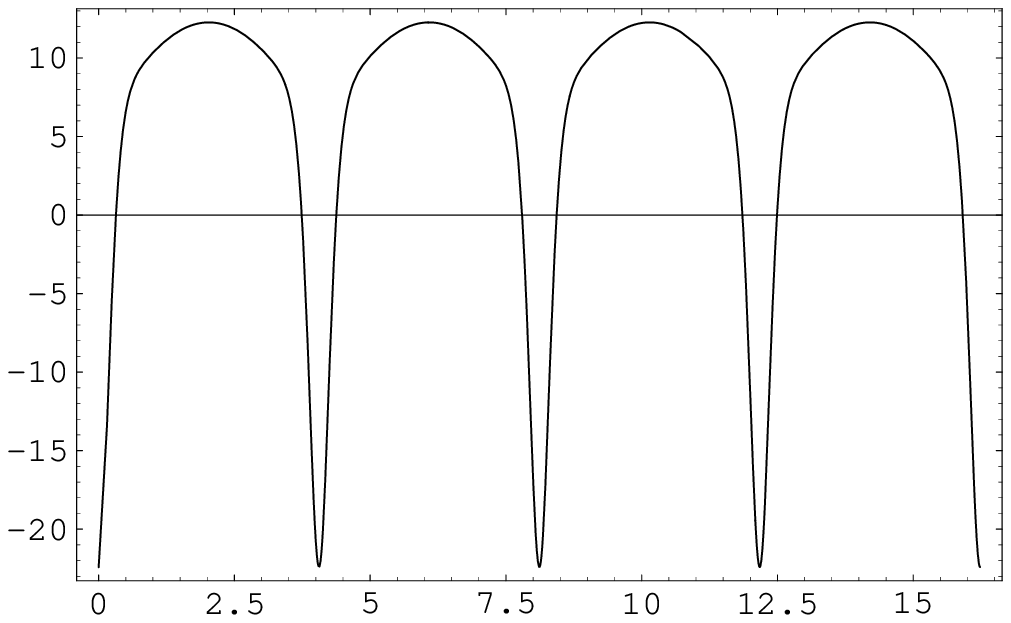,width=2.2in}\\ {\scriptsize{Fig. 1c$\qquad
m=0.66$}}
\end{tabular} &
\begin{tabular}{c}
\psfig{figure=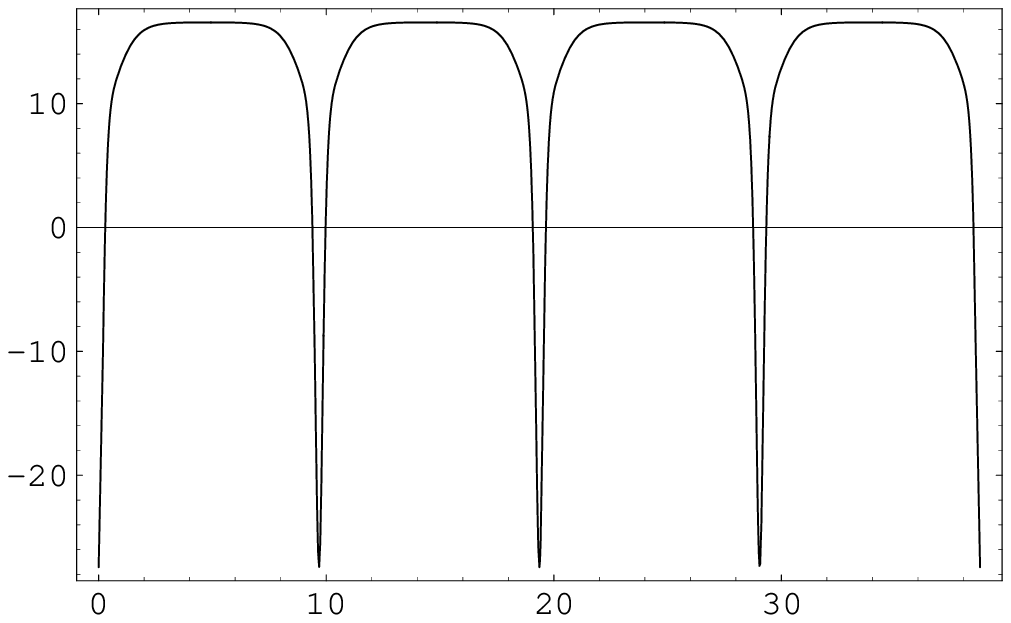,width=2.2in}\\ {\scriptsize{Fig. 1d$\qquad
m=0.999$}}
\end{tabular}
\end{tabular}
\end{center}
\caption{Different forms of potential \eqref{eq:x1lame} for
$\ell=4$ and various values of $m$. }\label{fig:x1potentials}
\end{figure}

Examples of $\mathrm{X}_1$ hyperbolic and polynomial potentials
and their square-integrable eigenfunctions have already been
presented in \cite{GKM3,GKM4}, and also in \cite{GT04,GT05} in the
context of supersymmetric quantum mechanics. Here we present an
example of a periodic $\mathrm{X}_1$ that can be regarded as a
modifcation of the well-known Lam\'e potential. The form of the
potential is
\begin{align}
  \label{eq:x1lame}
  u(x) &= \ m\, \ell(\ell+1)\, \sn(x| m)^2 +\\ \nonumber
  &\quad -
  \frac{12mb^2 +8(1-2m)b+4m-4}{\cn(x| m)^2-b} - \frac{8b(b-1)(mb-m+1)}{(\cn(x|
    m)^2-b)^2},
\end{align}
where $m\in (0,1)$, $b>2\ell/(2\ell-1)$ and where $\ell, m,
b$ are related by
\begin{equation}
  \label{eq:blmrel}
  (2\ell-1)^2=
  \frac{m (b-1)^2+2 b-1}{(b-1)^2\, b^2\, m}-\frac{m-1}{(bm-m+1)^2}.
\end{equation}
If $\ell$ is a positive integer, we set
\begin{equation*}
 k_1 = \frac{1}{4}-\frac{1}{2} \ell, \quad
 k_2 = \pm \frac{1}{4}
\end{equation*}
so that \eqref{eq:x1lame} becomes a specialization of
\eqref{eq:emodlamepot} and \eqref{eq:blmrel} becomes a
specialization of \eqref{eq:k1k2cond2}. A plot of the potential
for $\ell=4$ and various values of $m$ can be seen in Figure
\ref{fig:x1potentials}.

From \eqref{eq:k1k2gaugefactor}, the eigenfunctions have the
following form
\begin{equation*}
  \psi(x) = (\cn(x|m)^2 -b)^{-1} \dn(x|m)^{c_2} \cn(x|m)^{c_3}
  \sn(x|m)^{c_4} p(\cn(x|m)^2)
\end{equation*}
where $c_i = 2k_i + 1/2$ and $p(z)\in \cE_n^{(a,b)}(z)$ with $a$
given by \eqref{eq:k1k2aformula}. The potential has four algebraic
sectors. Depending on whether $\ell$ is even or odd, the four
possible algebraizations for each choice of $\ell$ are shown in
Table \ref{tab:x1lame}.
\begin{table}[htbp]
  \centering
  \begin{tabular}{cclll|clll}
    &\multispan{3}{\hfil $\ell$ even\hfil} &&     \multispan{4}{\hfil
      $\ell$ odd\hfil} \\
   Sector & $n$ & $c_2$ & $c_3$ & $c_4$  &    $n$ & $c_2$ & $c_3$ & $c_4$ \\
    \hline
    1&\vbox to 12pt{}$\ell/2$ & $1$ & $0$ & $0$ &  $\ell/2+1/2$ & $0$
    & $0$ & $0$ \\
    2&$\ell/2$ & $0$ & $1$ & $0$ &  $\ell/2-1/2$ & $1$ & $1$ & $0$ \\
    3&$\ell/2$ & $0$ & $0$ & $1$ &  $\ell/2-1/2$ & $1$ & $0$ & $1$ \\
    4&$\ell/2-1$ & $1$ & $1$ & $1$ &  $\ell/2-1/2$ & $0$ & $1$ & $1$  \\ \hline\\
  \end{tabular}
  \caption{The four algebraic sectors of the $\mathrm{X}_1$ elliptic potential \eqref{eq:x1lame}.}
  \label{tab:x1lame}
\end{table}

Having fixed $\ell$, the potential has only one remaining free
parameter, which we take to be $m$ and use \eqref{eq:blmrel} to
determine $b$, restricted to the condition $b>2\ell/(2\ell-1)$.
The free parameter $m$ ranges in the interval $(0,1)$. The two
limiting cases are
\begin{enumerate}
\item $m\to 0$, which implies that $b\to\infty$. The potential
\eqref{eq:x1lame} tends to a $\sin^2(x)$ trigonometric potential
(see Figure \ref{fig:x1potentials}a).

\item $m\to 1$, so that $b$ approaches $2\ell/(2\ell-1)$. The
potential in this limit assumes an interesting form of very deep,
short range wells on a uniform background (see Figure
\ref{fig:x1potentials}d).
\end{enumerate}

\begin{figure}[h]
\begin{center}
\begin{tabular}{cc}
\begin{tabular}{c}
\psfig{figure=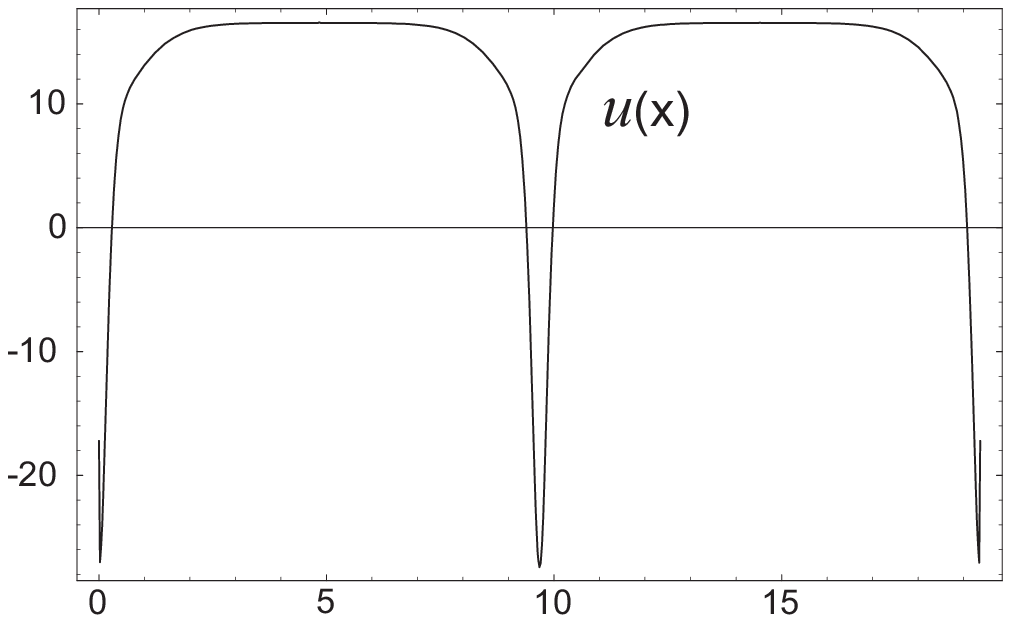,width=2.3in}\\{ \scriptsize{Periodic
potential $u(x)$  for $m=0.999$.}}
\end{tabular} &
\begin{tabular}{c}
\psfig{figure=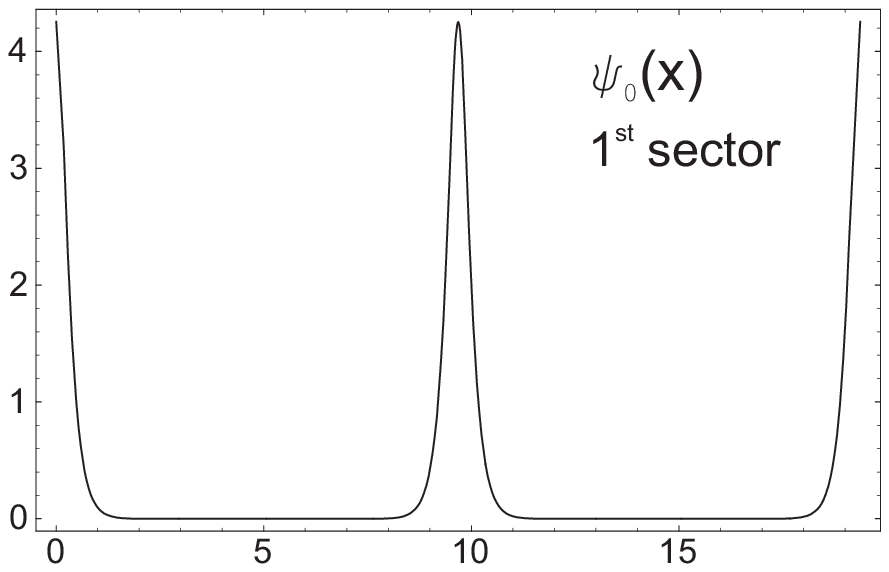,width=2.3in}\\ {\scriptsize{ $E_0=
-20.41235597893$} }
\end{tabular}\\\\
\begin{tabular}{c}
\psfig{figure=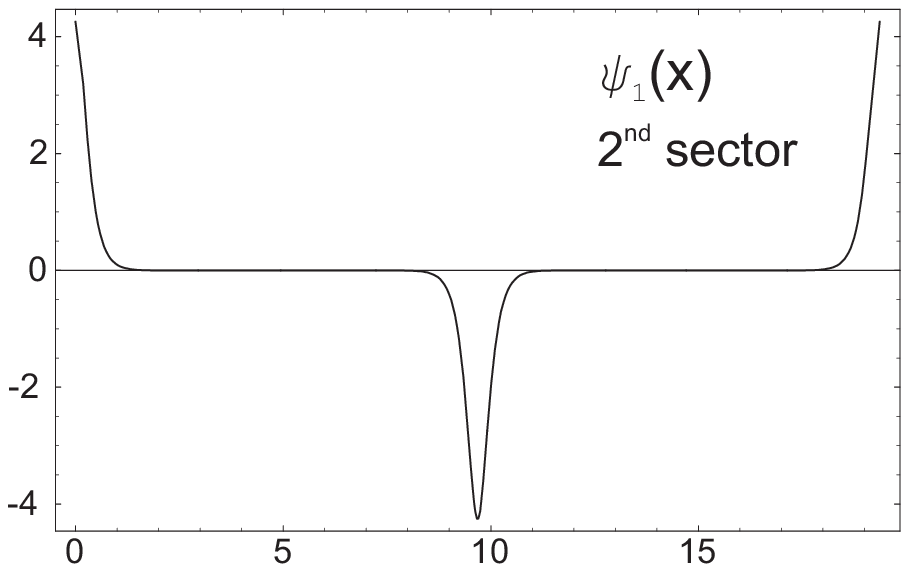,width=2.3in}\\ {\scriptsize{ $ E_1=
-20.41235597887$} }
\end{tabular} &
\begin{tabular}{c}
\psfig{figure=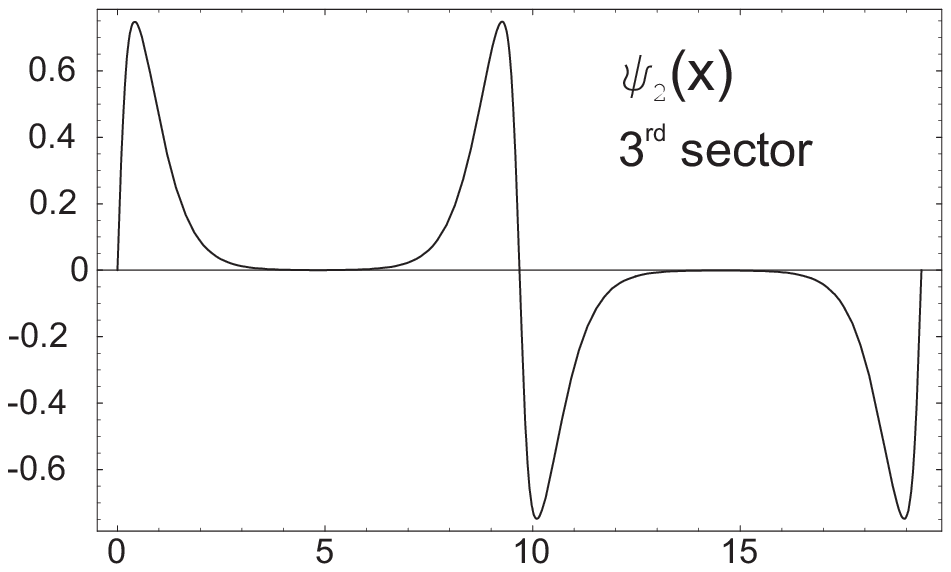,width=2.3in}\\ {\scriptsize{$ E_2=
0.577140437402$}}
\end{tabular}\\\\
\begin{tabular}{c}
\psfig{figure=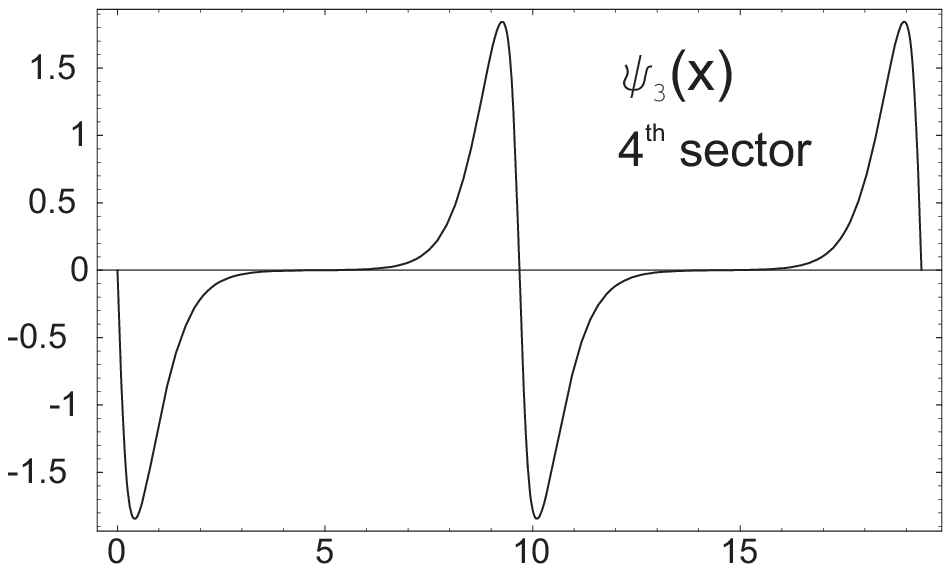,width=2.3in}\\ {\scriptsize{$ E_3=
0.577142188310$}}
\end{tabular} &
\begin{tabular}{c}
\psfig{figure=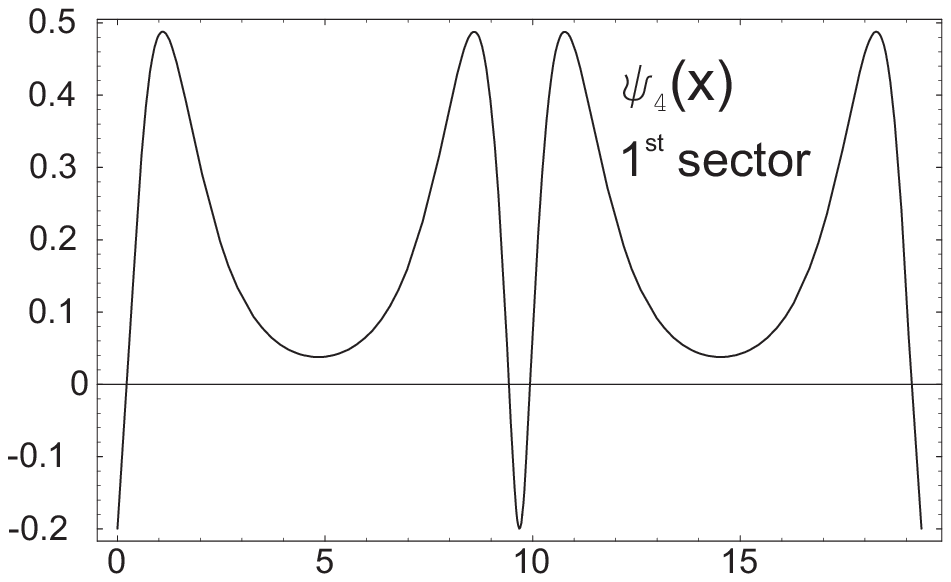,width=2.3in}\\ {\scriptsize{$ E_4=
3.57339250870$}}
\end{tabular}\\\\
\begin{tabular}{c}
\psfig{figure=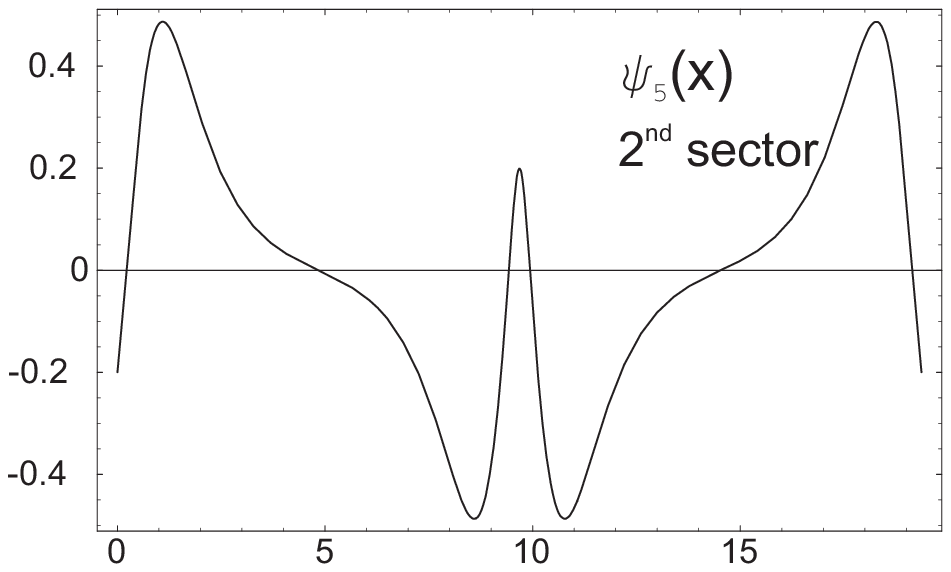,width=2.3in}\\ {\scriptsize{$ E_5=
3.57789250857$}}
\end{tabular} &
\begin{tabular}{c}
\psfig{figure=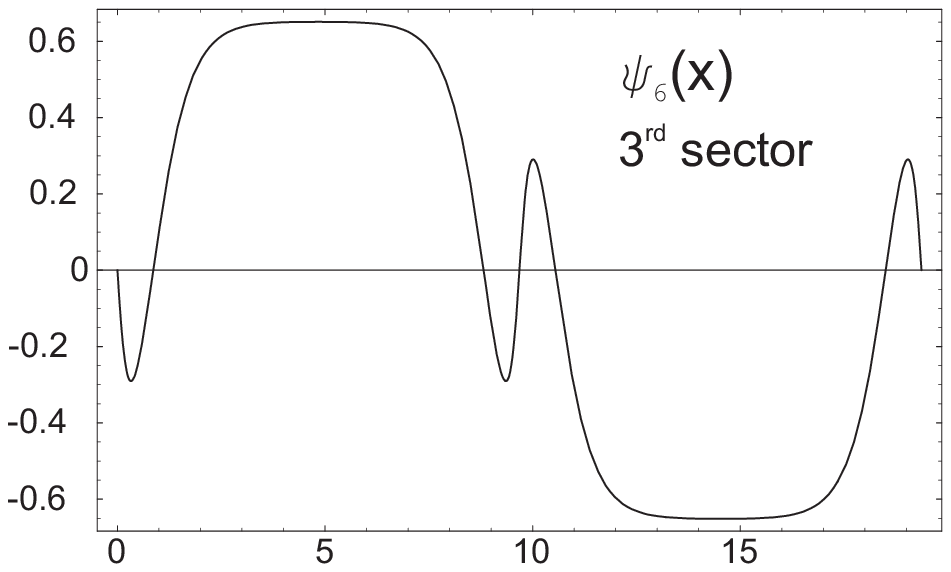,width=2.3in}\\ {\scriptsize{$ E_6=
4.57514431434$}}
\end{tabular}
\end{tabular}
\end{center}
\caption{The periodic potential $u(x)$ in \eqref{eq:x1lame} for
$\ell=4$ and $m=0.999$ has four gaps in its band spectrum. The
seven algebraic eigenfunctions that correspond to the band edges
are plotted together with their energies and the algebraic sector
to which they belong. }\label{fig:x1eigenfunctions}
\end{figure}

For integer values of $\ell$, potential \eqref{eq:x1lame} has
exactly $\ell-1$ gaps in its band spectrum  The $2\ell-1$
eigenvalues that describe the edges of the allowed and forbidden
energy bands (regions of stability or instability) can be
calculated algebraically. Note that Lam\'e potential
$m\ell(\ell+1)\sn^2(x|)$ has $\ell+1$ gaps in its band spectrum,
and the $2\ell+1$ eigenfunctions corresponding to the band edges
are the Lam\'e polynomials. The $\mathrm{X}_1$ elliptic finite gap
potential  \eqref{eq:x1lame} has one gap less than Lam\'e,
corresponding to the fact that the invariant polynomial space
$\cE_n^{(a,b)}(z)$ is a codimension one subspace of $\cP_n$.
For instance, when
$\ell=4$, the first three algebraic sectors are two-dimensional
while the fourth sector is one-dimensional (see Table
\ref{tab:x1lame}). For each of the four sectors, the actual
eigenvalues and eigenfunctions are calculated by diagonalizing the
matrix corresponding to the algebraic action shown in
\eqref{eq:k1k2T}. As an illustration, for $\ell=4$ and $m=0.999$
the seven algebraic eigenfunctions corresponding to the band edges
have been computed and displayed in Figure
\ref{fig:x1eigenfunctions}. A look at the energies shows that the
allowed bands are very narrow, which is natural since the value of
$m$ is very close to unity. In the limit $m\to1$ the period of the
potential diverges and the bands collapse into pure a point
spectrum.


\section{Summary and Outlook}\label{sec:summary}

In this paper we introduced the notion of an exceptional polynomial
subspace of $\cP_n$ and identified those subspaces as the only ones
that can give rise to new quasi-exactly solvable potentials in one
dimension. For exceptional subspaces of co-dimension one
($\mathrm{X}_1$ subspaces), we claimed (Theorem \ref{thm:barcelona})
that $\cE_n$ is the only $\mathrm X_1$ subspace up to an
$\mathrm{SL}_2$ transformation. Next, we used the equivalence
between an arbitrary second order differential operator and a
Schr\"odinger operator to construct, characterize and classify
$\mathrm X_1$ potentials, which are not Lie-algebraic. The
characterization is done at the level of the potential invariant,
and it involves an additional quadratic condition on the residues
$k_i$ of the quotient $q(z)/p(z)$ of the coefficients of the second
order operator. As in the Lie-algebraic case, the leading order
component $p(z)$ of an $\mathrm X_1$ operator is a quartic and we
have used $\mathrm{SL}_2$ covariance to classify them into canonical
forms, leading to new quasi-exactly solvable families of elliptic,
hyperbolic, trigonometric and rational potentials on the line. The
new characterization of the QES condition at the level of parameters
$k_i$ in the potential invariant allows to analyze the potentials
that admit multiple algebraic sectors, corresponding to a residual
symmetry in the choice of $k_i$ such that the potential remains
unchanged. We have provided a classification of all such cases.

Future work involves extending this classification into two
possible directions:
\begin{enumerate}
\item Exceptional subspaces of codimension two or higher. Some
examples of $\mathrm X_2$ subspaces are known to exist, but a full
classification is not yet available. The equivalence problem is
considerably harder than in the $\mathrm X_1$ case. \item
Exceptional subspaces of polynomials in two or more variables. A
full classification of exceptional Schr\"odinger in more than one
variable operators seems unfeasible due to the lack of a
constructive solution to the equivalence problem. Nevertheless, some
examples of many-body quasi-exactly solvable models are known
\cite{BTW98,FGGRZ00}, and we do not exclude that new many-body
problems exist with invariant spaces of exceptional type.
\end{enumerate}

\section*{Acknowledgments}

The research of DGU is supported in part by the Ram\'on y Cajal
program of the Ministerio de Ciencia y Tecnolog\'{i}a and by the
DGI under grants FIS2005-00752 and MTM2006-00478. The research of
NK and RM is supported in part by the NSERC grants RGPIN
105490-2004 and RGPIN-228057-2004, respectively.

%

\end{document}